# Time-Resolved Surface-Fault Displacement During the 2026 Kumamoto Earthquake From Near-Fault Video

Bogdan Enescu[1,2]*, Shinji Toda[3], Yuji Yagi[4], Kazuo Oike[5], Yu Xuan Teh[1], Shohei Nobuhara[6]

[1] Department of Geophysics, Graduate School of Science, Kyoto University, Kyoto, Japan

[2] National Institute for Earth Physics, Măgurele, Ilfov, Romania

[3] International Research Institute of Disaster Science, Tohoku University, Sendai, Japan

[4] Institute of Life and Environmental Sciences, University of Tsukuba, Tsukuba, Japan

[5] Global Center for Asian and Regional Research, University of Shizuoka, Shizuoka, Japan

[6] Faculty of Information and Human Sciences, Kyoto Institute of Technology, Kyoto, Japan

*Corresponding author: Bogdan Enescu (benescu@kugi.kyoto-u.ac.jp)

## Abstract

Direct observations of how displacement develops across a surface fault are rare, especially when a rupture has a large normal-fault component. The 28 July 2026 Kumamoto earthquake ($M_{JMA}$ 7.1; $M_w$ 6.8) produced approximately 33 km of mapped surface rupture along the Hinagu Fault Zone. Near Yatsushiro City, field measurements documented 1.05 m of right-lateral and 0.90 m of east-side-up displacement, giving a total vector of 1.383 m. A security camera at a convenience store close to the rupture recorded relative motion between the two fault blocks. The native camera file had been deleted automatically, and only a smartphone recording of the monitor survived. We measured image motion with two independent methods: Lucas–Kanade optical flow, which follows visible features, and normalized cross-correlation, which follows small image patches. After correcting for motion shared by reference structures on the camera side, the central 60% of the displacement developed in 0.866–0.901 s with optical flow and 0.910–0.928 s with cross-correlation. Scaling the histories to the field offset gives average rates of 0.920–0.958 and 0.895–0.912 m $s^{-1}$, respectively. A cross-check using independently defined image windows reproduces the displacement scale and rapid main rise. All analyses use the same secondary recording, so they share uncertainty in its frame timing and possible phone-to-screen distortion. The record also shows an early apparent peak and decline followed by renewed apparent motion, but it cannot establish a permanent endpoint or physical overshoot. The central 0.830 m of field-calibrated surface displacement developed in approximately 0.9 s at this oblique-slip rupture site.

## 1. Introduction

How slip develops during the few seconds in which an earthquake rupture passes a point is central to earthquake-source physics and near-fault hazard. Yet the second-by-second history of coseismic displacement at the ground surface is rarely observed directly. Seismic and geodetic data constrain rupture over broader spatial scales, whereas geological surveys ordinarily preserve only the final offset. Near-fault strong-motion instruments record rapid ground motion but usually sample one side of a fault and therefore do not directly measure relative displacement across the surface rupture. A time-resolved, across-fault observation can connect final geological offset to dynamic rupture, but such observations remain exceptional.

The 28 July 2026 Kumamoto earthquake provides such an opportunity. The $M_{JMA}$ 7.1 ($M_w$ 6.8) event occurred at 16 km depth in central Kyushu (Japan Meteorological Agency 2026; Figure 1). Rapid field investigation identified approximately 33 km of surface rupture along the Hinagu Fault Zone (Toda et al. 2026). Following the segmentation of the Headquarters for Earthquake Research Promotion (HERP 2013), Figure 1 labels the Hinagu and Takano-Shirahata segments. The Geospatial Information Authority of Japan (GSI 2026a) mapped a displacement boundary from Yatsushiro City toward Mifune Town, broadly following the fault zone but with local steps and bends in the Yatsushiro plain. At the video site, field measurements document 1.05 m of right-lateral and 0.90 m of east-side-up displacement. This large vertical component makes the local rupture strongly oblique, unlike the predominantly strike-slip surface rupture documented in recent video studies of the 2025 Myanmar earthquake (Kearse and Kaneko 2025; Latour et al. 2025; Hirano et al. 2025). Dipping-fault geometry can also produce asymmetric motion on the two sides of a rupture (Oglesby et al. 1998), making the local time history seismologically important.

Recent studies show that opportunistic video can provide this type of observation. Security-camera records of the 2025 $M_w$ 7.7 Myanmar earthquake have yielded local slip histories, trajectories, and rupture-speed constraints (Gao et al. 2025; Kearse and Kaneko 2025; Latour et al. 2025; Hirano et al. 2025). Hirano (2026) applied image correlation to records from the 2018 Hualien, 2025 Mandalay, and 2026 Kumamoto earthquakes, using event-specific fixed and tracked windows. Ando and Murata (2026) independently emphasized the conspicuous vertical component and an approximately 1-s displacement episode in the Kumamoto footage. These concurrent studies establish the value of near-fault imagery while also motivating careful separation of earthquake motion from camera and recording artifacts.

That separation is particularly difficult here because the native security-camera recording no longer exists. The circulated file was made by filming the security-camera monitor with a smartphone. Apparent motion may therefore come from shaking of the installed camera, movement of the phone, or changes in the angle and distance between the phone and screen. Display refresh, resampling, compression, and transcoding may add further distortion.

These effects cannot all be removed from the surviving file. Our common-motion correction removes only the part of the image motion shared by the near-side reference structures and described by the prespecified similarity or affine models. It does not explicitly remove moiré, display-refresh effects, rolling-shutter distortion, errors in frame timing, or a changing projective warp between the phone and monitor. Because every analysis uses the same secondary recording, agreement between trackers cannot independently verify its frame timing. We therefore use two trackers and fixed sensitivity tests to assess the large main rise, while treating the smaller late excursion as unresolved. The study asks how long the principal displacement took to develop, what average rate follows from the field offset, and which parts of the later image history cannot be interpreted physically.

## 2. Earthquake and observational setting

### *2.1. Surface rupture and field displacement*

The observation site is beside a convenience store in Miyaji, Yatsushiro City (Figures 1 and 2a). The mapped GSI displacement boundary bends through the local road and paddy-field setting. A field photograph shows sharp deformation at a paddy margin and damage to an adjacent concrete path (Figure 2b). We call the block containing the security camera the camera-side (near-side) block and the block across the mapped rupture the opposite-side (far-side) block. We use the shorter near-side and far-side terms below.

Field measurements corresponding to Photo 17 of the preliminary survey give 1.05 m of right-lateral displacement and 0.90 m of east-side-up displacement (Toda et al. 2026). The field team estimates an approximate ±0.05 m tape-measurement uncertainty for each component. Their vector magnitude is $D_f = \sqrt{(1.05^2 + 0.90^2)} = 1.383$ m. Treating the component uncertainties as independent gives an approximate ±0.050 m uncertainty in $D_f$ (±3.6%). We use this independently measured magnitude for a scalar image-to-ground calibration and report its uncertainty separately from image-processing sensitivity. Because neither the original camera position nor the full three-dimensional viewing geometry was surveyed, we do not convert image axes separately into geological strike-slip and dip-slip components.

### *2.2. Video source and timing*

The analyzed source is the publicly circulated smartphone recording of the security-camera monitor. According to information obtained from the video provider through Yomiuri Shimbun, the store system deleted the native recording automatically after ten days, and it could not be recovered. The copy contains 442 frames, has a resolution of 1276 × 720 pixels, and lasts 14.733 s. Its decoded frames carry uniformly spaced timestamps at a nominal rate of 30 frames $s^{-1}$; we use these timestamps as elapsed time. This is an assumption about the circulated file because the original camera timing, monitor refresh, and phone-capture timing cannot be reconstructed. All analyses use this same decoded file without spatial resizing (Text S1).

The predefined primary analysis spans video time 5.0–11.0 s (181 frames), including a 5.0–6.5-s pre-displacement baseline. An exploratory diagnostic extends the histories through the last usable frame at 14.700 s. All times in this paper are measured from the beginning of the supplied secondary video; they are not times relative to the occurrence of the earthquake.

### *2.3. Image geometry*

The primary target region of interest (ROI) contains buildings and other textured structures assigned to the far-side block. Separate reference ROIs contain structures assigned to the near-side block (Figure 2c). The broad target ROI covers many stable structures; the paddy itself has little visual texture and contributes few tracked features. The block assignment is supported by the mapped GSI boundary, field observations, local perspective, and viewing direction. It is also consistent with the geometry used independently by Hirano (2026): two fixed windows on the near-side block and one tracked window on the far-side block. Hirano supplied the exact coordinates of these three 50 × 50-pixel windows. We use them only as a restricted-geometry cross-check, not as replacements for the primary ROIs.

To show the surface break in the oblique camera view, we aligned a post-motion Yomiuri still with the 5.0-s pre-slip frame and transferred only the clearly identifiable rupture-zone segments (Figure 2c; Text S2). The alignment used 19 geometrically consistent image matches and has a root-mean-square residual of 1.37 pixels. Because 17 of the 19 matches fall inside the far-side target ROI, the alignment is well constrained only in that part of the frame. It does not independently determine where the rupture would project near the near-side reference ROIs. We therefore display only the locally supported segments and do not extend them toward the reference ROIs. The overlay provides visual context only; it was not used to select tracking features or calculate displacement.

## 3. Measuring displacement from the video

### *3.1. Two independent tracking methods*

We used two OpenCV-based tracking methods (Bradski 2000) that measure image motion in different ways. Pyramidal Lucas–Kanade optical flow (Lucas and Kanade 1981) follows distinct visible features from the 5.0-s baseline frame directly to each later frame. The features are selected with the Shi–Tomasi method (Shi and Tomasi 1994), and a forward–backward check rejects features that cannot be followed consistently. Normalized cross-correlation (NCC) instead follows small fixed image patches by finding where each patch reappears in a later frame. We retained three patch sizes. Texts S3 and S4 and Tables S1 and S2 give the complete selection, tracking, and acceptance settings.

### *3.2. Correcting motion shared across the recording*

The near-side reference features contain recording motion but should not contain the relative displacement across the fault. For each frame, random sample consensus (RANSAC; Fischler and Bolles, 1981) kept reference tracks that moved consistently and rejected outliers. We described the shared motion with two models chosen before examining the result. A similarity transform allows translation, rotation, and a uniform change in scale. An affine transform also allows different scale changes in different directions and shear. Optical flow used two ways of weighting the reference regions (feature-weighted and patch-balanced), each combined with the two motion models, giving four variants. NCC used three patch sizes (31, 41, and 51 pixels), each combined with the two motion models, giving six variants. Texts S3 and S4 list the fixed RANSAC settings.

For each far-side target feature, the fitted reference model estimates where that feature would appear if it moved only with the near-side reference structures. The difference between its observed and predicted positions is the relative image displacement across the fault. We represent the target motion in each frame by the two-dimensional median of the accepted residuals. This correction removes only motion represented by the reference-derived similarity or affine model. We also tested a projective transformation, or homography, which can describe changes in perspective when a phone moves relative to a flat screen. Those tests were used only to diagnose recording artifacts because they did not pass validation; no homography correction was applied to the primary displacement history.

### *3.3. Measuring rise time and scaling to the field offset*

We smoothed the horizontal and upward histories with a second-order Savitzky–Golay filter (Savitzky and Golay 1964). Seven windows, fixed in advance, range from 5 to 17 frames (0.17–0.57 s at 30 frames $s^{-1}$). Six prespecified 0.5-s intervals between 10.0 and 11.0 s provide alternative reference levels for the end of the main rise. For each combination of tracker settings, shared-motion model, smoothing window, and reference interval, we remove the 5.0–6.5-s baseline, recompute the two-dimensional displacement direction, project the motion onto that direction, and normalize it by the interval median. These intervals are analysis reference levels, not assumed settled plateaus (Text S5).

Our principal timescale is the time required for displacement to rise from 20% to 80% of the chosen reference level. We denote it $T_{20-80} = t_{80} - t_{20}$ and interpolate the first upward crossings between adjacent frames. This central 60% interval avoids the low-amplitude beginning and end of the motion, where noise and recording artifacts have greater influence. We did not choose one smoothing window after viewing the result; all seven are retained in the primary analysis. The 10–90% duration and instantaneous peak rate are secondary quantities.

To express the displacement in metres, we match the image-vector magnitude $|D_{px}|$ for each reference interval to the measured field-vector magnitude $D_f = 1.383$ m. This is an empirical one-factor calibration, not a reconstruction of the three-dimensional camera geometry. Within each processing realization, the same scale factor is used throughout the time history. This assumes that the effective image scale does not change materially during the main rise; movement of the phone relative to the monitor could violate that assumption. The calibration assigns the 1.383-m field offset to the chosen image endpoint. The 20–80% interval therefore always represents 60% of 1.383 m, or 0.829759 m, and its average rate is $V_{20–80} = 0.829759$ m/$T_{20–80}$. The endpoint's pixel magnitude consequently does not appear in this rate formula. Choosing a different endpoint interval can still change the projected direction and the 20% and 80% crossing times, so we recompute those quantities for every endpoint test rather than simply rescaling one curve.

### *3.4. Prespecified sensitivity tests and validation*

The ranges in Table 1 show how the result changes across analysis choices fixed before the tests; they are not statistical confidence intervals. They retain every predefined tracker variant, smoothing window, and endpoint interval. We separately tested smoothing and other processing choices (Texts S6 and S7), reference-region selection (Texts S8–S10), target-region location and size (Text S11), and endpoint definition (Text S12). These ranges do not include possible biases shared by every analysis of the same secondary recording, such as errors in its frame timing or uncorrected phone-to-monitor projective distortion. The field team's approximate ±0.05 m uncertainty for each measured component corresponds to about ±3.6% uncertainty in the scalar calibration and calibrated rates and is reported separately.

A geometry check used the two near-side windows and one far-side window used by Hirano (2026), at the exact coordinates supplied by Hirano. We report the number of tracked features, their frame-by-frame support, and the NCC variants that failed because the windows contained too few templates (Texts S13 and S14). Other checks omitted reference regions one at a time, shifted or resized the target region, examined the display border, and tested projective corrections (Text S15). We did not change a threshold, smoothing window, motion model, normalization interval, or acceptance rule to obtain a preferred timing or overshoot result.

## 4. Results

### *4.1. Principal surface-displacement history*

Both trackers resolve the same rapid, multi-stage main rise (Figure 3; Figure S1). The median histories briefly flatten near video time 7.9–8.0 s, at about 35–40% of the selected endpoint, before the steeper rise resumes. This shoulder is visible in both tracker ensembles, but the 30-frame-$s^{-1}$ sampling, smoothing dependence, and common secondary source do not justify interpreting it as a separate physical slip pulse. Across the primary grid, $T_{20–80}$ is 0.866–0.901 s for optical flow and 0.910–0.928 s for NCC (Table 1; Table S3). The ranges do not overlap, so we report them separately, but both imply a central timescale of approximately 0.9 s.

The image endpoint components are 18.394 pixels horizontal and 15.811 pixels upward for optical flow, and 18.929 and 16.086 pixels for NCC. The horizontal/upward image ratios are therefore 1.163 and 1.177. These values bracket the corresponding ratio of the field-measured right-lateral and east-side-up components, 1.05/0.90 = 1.167. This descriptive agreement indicates that the field displacement vector is viewed approximately in the plane of the image, but it is not a three-dimensional reconstruction. Median vector magnitudes are 24.280 pixels [23.996–24.389] and 24.844 pixels [24.758–24.948]. Matching these magnitudes to the 1.383-m field vector gives scalar image-to-ground conversion factors of 5.696 cm $pixel^{-1}$ [5.670–5.763] and 5.566 cm $pixel^{-1}$ [5.543–5.586]. Component and magnitude medians were aggregated separately and therefore need not recombine exactly.

### *4.2. Field-calibrated displacement rate*

The central 60% of the measured field vector is 0.829759 m. Dividing by the unrounded rise times gives average rates of 0.920–0.958 m $s^{-1}$ for optical flow and 0.895–0.912 m $s^{-1}$ for NCC. Rates and displayed durations are rounded independently. Instantaneous peak rates are higher—median values of 1.428 and 1.275 m $s^{-1}$—but vary appreciably with smoothing and are therefore secondary results (Table S4; Figures S2 and S3).

### *4.3. Sensitivity to region and motion-model choices*

Figure 4 compares the main sources of processing sensitivity. For NCC, the choice between similarity and affine correction has the largest effect, spanning 0.068 s. Across the complete set of smoothing windows, the method median changes by 0.044 s for optical flow and 0.036 s for NCC, about 5% and 4% of the approximately 0.9-s duration. Smoothing therefore affects the precise duration but does not create the main rise. Endpoint normalization spans 0.018–0.035 s, and shifting or resizing the target ROI has median effects of 0.005–0.008 s. Longer smoothing windows shorten $T_{20–80}$ here because they round small shoulders near the 20% and 80% levels and move the first crossings inward; this is a measurement effect, not faster ground motion. Figure 4a holds the smoothing window fixed, whereas Table 1 combines all seven windows, so the displayed medians need not have identical ranges.

Reference-region omission tests show that no single reference zone creates the early peak-and-decline pattern (Figure S4). The pattern satisfies the internal video-history classification in most frozen realizations (Table S5; Figure S5), but this is not a physical-overshoot test. Endpoint and target-region tests show that the approximately 0.9-s main rise is more stable than the few-percent early excess (Tables S6 and S7; Figure S6).

The geometry check using the window coordinates supplied by Hirano retained 29 near-side reference features and 16 far-side target features; optical flow remained valid in all 181 frames (Table S8; Figure S7). Both shared-motion models recovered the rapid main rise: similarity correction gives $T_{20–80}$ = 0.826–0.852 s and affine correction gives 0.842–0.892 s (Figure S8). The primary multi-template NCC method could not be used in these small windows because they contained too few templates. As a separate check, tracking each complete 50 × 50-pixel window gives 0.826–0.846 s and reproduces the endpoint magnitude to within 0.11% (Figure S9). Because all of these checks use the same recording, they test geometry and processing behavior but are not independent observations.

Table S9 summarizes attempts to identify artifacts in the secondary recording. Neither the scene-reference homography nor the display-plane homography passed its validation tests, so neither was applied to the earthquake history (Figure S10). An exploratory row-wise stabilization reduces large apparent image motion, but its soft constraint overlaps 57.9% of the far-side target ROI and could remove real across-fault motion. We therefore retain it only as a diagnostic (Text S15). These tests show that a more flexible correction is not automatically safer when it cannot be shown to preserve the earthquake signal.

Figure S12 summarizes the full workflow, and Table S10 reports the similarity- and affine-specific timing results. The diagnostics limit interpretation of the late history but do not alter the primary displacement curve.

### *4.4. Late apparent motion*

Relative to the six primary normalization intervals, the early local displacement excess is 4.640–6.667% for optical flow and 4.241–5.491% for NCC. The extended histories do not settle at the original level (Figure 5). The exploratory 14.033–14.533-s level is 7.880% above the original optical-flow level and 6.325% above the original NCC level; both late histories rise above the earlier apparent peak. Recomputing direction and crossings with this late level gives exploratory $T_{20\text{–}80}$ values of 0.972 and 0.998 s, corresponding to 0.853 and 0.831 m s$^{-1}$. These values are diagnostics rather than replacements for the primary grid.

Artifact controls do not resolve the origin of the late motion. Reference-only projective correction retains 58% of the optical-flow late increment but only 4% of the NCC increment, while held-out reference validation fails for both. The display-border control lacks horizontal observability and stable late tracking. Whole-window NCC does not pass the frozen >3σ gate for the small early-excess feature. The defensible observation is therefore an early apparent peak and decline followed by renewed apparent horizontal displacement; the permanent endpoint and physical overshoot remain unresolved.

## 5. Discussion

### *5.1. Kinematics of an oblique-slip surface rupture*

The principal seismological result is that 0.830 m—the central 60% of the 1.383-m field vector—developed within approximately 0.9 s. The corresponding average rate is close to 0.9 m s$^{-1}$. The 1.05-m right-lateral component records strike-slip motion. Together with the northwestward dip of the fault, the 0.90-m east-side-up component indicates a substantial normal-slip component (HERP 2013; Toda et al. 2026). The similar image and field component ratios indicate that the displacement direction is viewed approximately in the image plane. Nevertheless, we use only the total field-vector magnitude to calibrate the image history; we do not separately convert image-horizontal and image-vertical motion into geological slip components.

The observation adds a time constraint to the mapped surface rupture. Field and GSI observations show where the ground broke and how much permanent displacement remained; the video shows that most of the local offset accumulated in roughly one second. This information is useful for dynamic-rupture models of dipping and oblique-slip faults, where fault geometry can make motion asymmetric across the rupture (Oglesby et al. 1998). Recovering the fault-plane slip history would require surveyed camera geometry, a native recording, and independent strong-motion or geodetic constraints.

The video measures relative displacement between the two fault blocks, not the absolute ground acceleration at either location. It therefore cannot be used to estimate the inertial loads on nearby buildings. What it does show is that a structure crossing the rupture would have experienced about 0.83 m of differential ground movement during the central 0.9-s rise. This rapid permanent deformation is a direct hazard for roads, foundations, pipelines, and other facilities that cross the fault, distinct from the hazard caused by ground shaking.

### *5.2. Relation to other video observations*

For the 2025 $M_w$ 7.7 Myanmar earthquake, Kearse and Kaneko (2025) reported 2.5 ± 0.5 m of displacement over 1.3 ± 0.2 s and a peak velocity of 3.2 ± 1.0 m $s^{-1}$, while Latour et al. (2025) reported approximately 3 m, a local duration near 1.4 s, and a peak velocity near 3.5 m $s^{-1}$. Kumamoto has a smaller 1.383-m field offset. Using the same instantaneous-peak definition, the 11-frame-smoothed Kumamoto histories give 1.428 m $s^{-1}$ for optical flow and 1.275 m $s^{-1}$ for NCC, lower than the Myanmar estimates. The Kumamoto peaks are smoothing-sensitive: the complete fixed-grid ranges are 1.192–1.712 m $s^{-1}$ for optical flow and 0.961–2.469 m $s^{-1}$ for NCC. Thus, the peak-rate and offset contrast is descriptive rather than a scaling relation, because the events also differ in mechanism, site, viewing geometry, recording provenance, and processing definition. In both events, most of the meter-scale surface displacement developed over about one second.

Hirano (2026) and Ando and Murata (2026) also recover rapid Kumamoto displacement from the same event and recording. Hirano (2026) emphasizes that the Kumamoto velocity history is more complex than those of the other two events he examined. The shoulder visible in both of our tracker medians is consistent with a complex image history, but we do not interpret it as a specific number of physical slip pulses. Our study adds a prespecified ensemble of two trackers, calibration to the field displacement vector, separate results for the similarity and affine corrections, and a record of failed diagnostics. Agreement among analyses of the same recording is useful, but it is not independent evidence and all analyses share the limitations of the circulated source.

### *5.3. What the validation does and does not establish*

Both trackers recover the approximately 0.9-s main rise, and reasonable changes in endpoint and target region have little effect. This agreement shows that the main rise is not specific to one tracking method, but it cannot test errors in frame timing or geometric distortion shared by the same source video. Within the primary NCC analysis, the choice of shared-motion model remains the most influential processing decision. Affine correction can describe directional scale change and shear, whereas similarity correction cannot. The secondary recording and available reference geometry do not show which is physically preferable, so we report the two NCC model groups separately instead of selecting one after seeing the timing. In the test using Hirano's supplied window coordinates, the similarity and affine ranges overlap; the much larger split obtained earlier from approximately digitized windows is not reproduced and is superseded.

The early peak and decline in Figure 5 cannot be identified as physical overshoot. Dynamic overshoot and stopping phases can be interpreted when seismic, geodetic, or original-quality imaging data resolve the source history (Ide et al. 2011; Yagi et al. 2025; Kearse and Kaneko 2026). Here, the late level lies above both the original normalization interval and the earlier apparent peak. If the late rise is continued ground displacement, the original interval underestimates the permanent level. If it comes from changing phone-to-screen geometry or another recording artifact, replacing the original level with the late level would also be unjustified. Strong-motion records could test whether late ground motion occurred, but they cannot determine whether this particular image excursion is instrumental.

### *5.4. Recording provenance, limitations, and data preservation*

Automatic deletion of the native file is the main limitation. Field measurements set the displacement scale, and the two trackers constrain the main rise. Spatial tests and Hirano's windows check the near-side/far-side assignment within the same recording. None of these tests independently validates the secondary video's timing or removes every phone-to-monitor distortion, and the late-history controls remain inconclusive. Reanalysis of the circulated copy cannot recover the original camera metadata, monitor refresh, phone motion, or earlier encoding steps.

This case shows why earthquake videos should be preserved quickly. Native security-camera files should be exported before automatic overwrite, with their original metadata and timing. Camera mounting, viewing direction, lens and stabilization settings, display geometry when relevant, and independent nearby imagery should also be documented. Such records could then provide stronger quantitative constraints on surface-rupture kinematics.

## 6. Conclusions

A near-fault security camera recorded rapid relative motion across the surface rupture during the 28 July 2026 Kumamoto earthquake. Field observations at the site document 1.05 m of right-lateral and 0.90 m of east-side-up displacement. Optical flow places the central 60% rise at 0.866–0.901 s and NCC at 0.910–0.928 s. Calibration to the 1.383-m field vector gives average rates of 0.920–0.958 and 0.895–0.912 m $s^{-1}$.

The principal result is that 0.830 m, the central 60% of the measured field displacement, developed in approximately 0.9 s at an oblique-slip rupture site with a substantial normal component. Across the full set of smoothing windows, the estimated duration changes by no more than 0.044 s for optical flow and 0.036 s for NCC. The cross-check using Hirano's supplied window coordinates also reproduces the rapid rise and displacement scale. Both tracker ensembles show a shoulder during the rise, but the recording does not establish separate physical slip pulses.

The video constrains rapid relative displacement across the fault, but it does not measure absolute ground acceleration or the resulting forces on nearby structures. The later image history also remains ambiguous. It contains an early apparent peak and decline followed by renewed apparent motion, but the secondary recording cannot establish a settled permanent endpoint or physical overshoot. These limits should remain part of the scientific conclusion.

## List of abbreviations

CCTV: closed-circuit television; GSI: Geospatial Information Authority of Japan; HERP: Headquarters for Earthquake Research Promotion; JMA: Japan Meteorological Agency; NCC: normalized cross-correlation; RGAFJ: Research Group for Active Faults of Japan; ROI: region of interest; RANSAC: random sample consensus.

## Declarations

### *Ethics approval and consent to participate*

Not applicable.

### *Consent for publication*

All displayed video frames and the field photograph were privacy-protected so that no identifiable personal information is presented. The analyzed recording was publicly available through a Threads post (Threads 2026). Only limited still frames are reproduced for scholarly analysis and explanation. The authors believe that this use is consistent with the copyright limitations and exceptions applicable to scholarship, research, and quotation. Copyright in the source recording and reproduced frames remains with the respective rights holder. The video-derived images are excluded from the article's Creative Commons Attribution 4.0 licence, and permission from the rights holder is required for further reuse.

### *Availability of data and materials*

The analyzed video is the publicly circulated secondary recording of the security-camera monitor (SHA-256 F3DD985B4D31EF5DBE77822C46E084A088149E5AF098557F2954676FDF73FB51). The native recording was automatically deleted and could not be recovered. The complete reproducibility archive contains the scripts, frozen configurations, derived histories, normalization outputs, region definitions, feature classifications, frame-by-frame support counts, registration controls, and records of failed or inconclusive diagnostics reported in this paper. It is publicly available at Zenodo (https://doi.org/10.5281/zenodo.22799211). The archive does not redistribute the source video; the checksum above identifies the exact circulated file used in the analyses.

***Competing interests***

The authors declare that they have no competing interests.

***Funding***

B.E. acknowledges support from JSPS KAKENHI Grant Number JP26K07241. The funder had no role in study design, analysis, interpretation, or preparation of the manuscript.

***Authors' contributions***

B.E. conceived the study, performed the primary image analysis and sensitivity tests, and drafted the manuscript. S.T. contributed field observations and geological interpretation. Y.Y. and K.O. contributed seismological interpretation and manuscript revision. Y.X.T. contributed independent validation and reproducibility checks and manuscript review. S.N. contributed computer-vision methodology, designed and interpreted the display-plane stabilization diagnostics, and revised the manuscript. All authors reviewed and approved the manuscript.

***Acknowledgements***

We thank the video owner and Yomiuri Shimbun staff for clarifying the recording provenance. We thank the field-survey team for rapid documentation of the surface rupture and permanent displacement. We acknowledge the Geospatial Information Authority of Japan for the seamless aerial imagery and georeferenced displacement-boundary data. We thank T. Takeda for providing the digital mapped-fault file used in Figure 1; its source is the Research Group for Active Faults of Japan (1991).

***Use of generative artificial intelligence***

OpenAI ChatGPT and Codex, as well as Anthropic Claude, were used during manuscript preparation to assist with language editing, improving clarity and consistency, and the development and checking of portions of the image-analysis and sensitivity-testing code. All analyses were executed and verified by the authors, and all methodological choices, scientific interpretations, and conclusions were reviewed and approved by the authors.

**Additional file**

Additional file 1. Supporting Information (PDF). Supplementary methods, exact processing settings, validation results, Texts S1–S15, Tables S1–S10, and Figures S1–

S12.

**Tables**

**Table 1.** Primary timing and rate ranges, with the exploratory late-endpoint diagnostic

| **Method** | **Primary $T_{20–80}$ (s)** | **Primary rate (m s⁻¹)** | **Original-interval median [p16–p84] (s)** | **Late $T_{20–80}$ (s)** | **Late rate (m s⁻¹)** |
|---|---|---|---|---|---|
| Optical flow | 0.866–0.901 | 0.920–0.958 | 0.867 [0.850–0.894] | 0.972 | 0.853 |
| NCC template matching | 0.910–0.928 | 0.895–0.912 | 0.910 [0.863–0.943] | 0.998 | 0.831 |

Note. Primary ranges are ranges of method medians over the six prespecified endpoint-normalization intervals, with all predefined tracker variants and smoothing windows retained. Original-interval values are medians [16th–84th percentile processing spreads] for 10.0–10.5 s. The late endpoint is exploratory. Rates were calculated from unrounded $T_{20–80}$ values; displayed durations and rates are rounded independently. The field team's approximate ±0.05 m uncertainty for each measured displacement component corresponds to about ±3.6% uncertainty in the scalar calibration and calibrated rates and is not included in these processing ranges. The ranges also do not include possible timing or geometric biases shared by every analysis of the secondary recording.

**Figure legends**

Figure 1. Regional seismotectonic setting. (a) Mainshock hypocenter and depth from JMA, the 2026 GSI surface-displacement boundary, mapped active-fault traces from the Research Group for Active Faults of Japan (RGAFJ, 1991), and the near-fault video site in the Yatsushiro-Mifune region. The mapped active-fault traces show approximate regional fault distributions and are distinct from the 2026 coseismic displacement boundary shown in red. The Hinagu and Takano-Shirahata segments of the Hinagu Fault Zone are labeled following HERP (2013). (b) Location of the study area in Japan. The site symbol points to the local view shown in Figure 2.

Figure 2. Geological setting, field evidence, and image-analysis geometry. (a) GSI seamless aerial imagery showing the approximate observation/store site and the geospatially projected GSI displacement boundary. The field displacement measured by Toda et al. (2026) was 1.05 m right-lateral and 0.90 m east-side-up; the field team estimates an approximate ±0.05 m tape-measurement uncertainty for each component. (b) Privacy-protected field photograph showing the rupture beside the paddy and damage to the adjacent concrete path. The red line and arrows show the field interpretation supplied by S. Toda; they are visual guides, not surveyed geometry. Photo: S. Toda. (c) Representative pre-slip frame at video time 5.0 s. Blue boxes are primary near-side reference ROIs; the orange box is the primary far-side target ROI. Dotted green and magenta boxes are the 50 × 50-pixel windows defined by coordinates supplied by Hirano and corresponding to his published geometry and interpreted here as near-side and far-side, respectively. Registered rupture-zone segments are shown only where they can be transferred reliably from the post-motion Yomiuri still and are not extrapolated across the reference regions. The near-side/far-side block assignments are instead supported by the mapped GSI displacement boundary, the local perspective and approximate viewing direction, the field observations and photographs of Toda et al. (2026), and their consistency with the fixed/reference and tracked/target window geometry published by Hirano (2026). A projective homography uses 19 geometrically consistent registration inliers, 17 of which lie within the primary far-side target ROI (Text S2). The segments are shown for geometrical context and were not used algorithmically to choose or reject tracking features. The region containing the lower-right parked vehicle was excluded from the reference area (Text S1). Privacy blurring affects display only. Video frame from the public Threads post cited in the text; excluded from the article's CC BY 4.0 licence.

Figure 3. Primary displacement histories from optical flow and NCC. Optical flow is shown by a solid blue line and NCC by a dashed orange line. (a) Displacement projected onto the interval-defined direction and normalized to the original 10.0–10.5-s interval; shaded envelopes show the 16th–84th percentile tracker-variant spread at 11-frame smoothing. In both panels, dotted horizontal lines mark the 20% and 80% crossing levels; the dashed horizontal line marks the selected endpoint (1.0 in panel a and 1.383 m in panel b). Both method medians show a shoulder near video time 7.9–8.0 s, at approximately 35–40% of the selected endpoint, before the steeper rise resumes. (b) The same histories after scalar calibration to $D_f = 1.383$ m. Within each realization, one scale factor is held fixed throughout the history. Across the endpoint-normalization grid, $T_{20–80}$ is 0.866–0.901 s for optical flow and 0.910–0.928 s for NCC.

**Figure 4.** Primary timing robustness across prespecified processing choices. Optical flow is blue and NCC template matching is orange. (a) Method-median $T_{20–80}$ as a function of the seven fixed Savitzky–Golay smoothing windows for the original 10.0–10.5-s normalization interval; shading gives the 16th–84th percentile spread across tracker variants. Panel (a) conditions on one smoothing window, whereas Table 1 pools all seven, so its smoothing-specific medians can extend beyond the headline ranges. (b) Method-median $T_{20–80}$ across the six fixed 0.5-s normalization intervals, with endpoint direction and threshold crossings recomputed for each interval; shading gives the 16th–84th percentile processing spread. (c) Minimum-to-maximum range of the endpoint-median $T_{20–80}$ values for similarity and affine common-motion correction across the six normalization intervals. These sensitivities are reported separately and were not retuned or collapsed into a single uncertainty estimate.

Figure 5. Late displacement history and endpoint sensitivity. Optical flow is solid blue and NCC is dashed orange. Both panels show dimensionless apparent displacement normalized to the stated interval; no pixel-to-metre conversion is applied in this figure. (a) Early history normalized to the original 10.0–10.5-s interval, showing the apparent peak and decline; the grey band marks that normalization interval. (b) History through 14.700 s, showing renewed apparent displacement; the grey and red bands mark the original and exploratory 14.033–14.533-s intervals. Lines are method medians and shading gives the 16th–84th percentile processing spread.

# Figures

## Figure 1

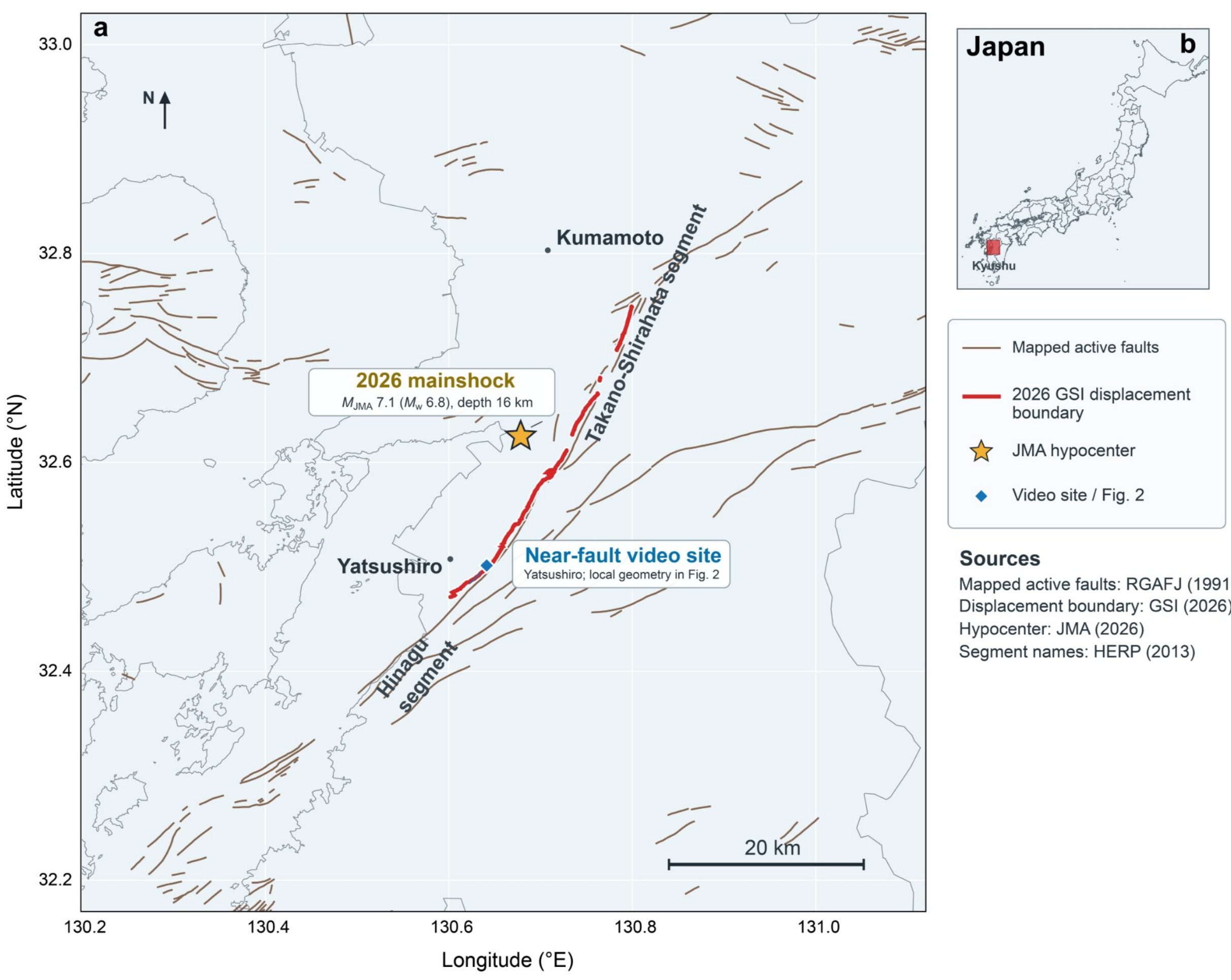

## Figure 2

Site and mapped displacement boundary

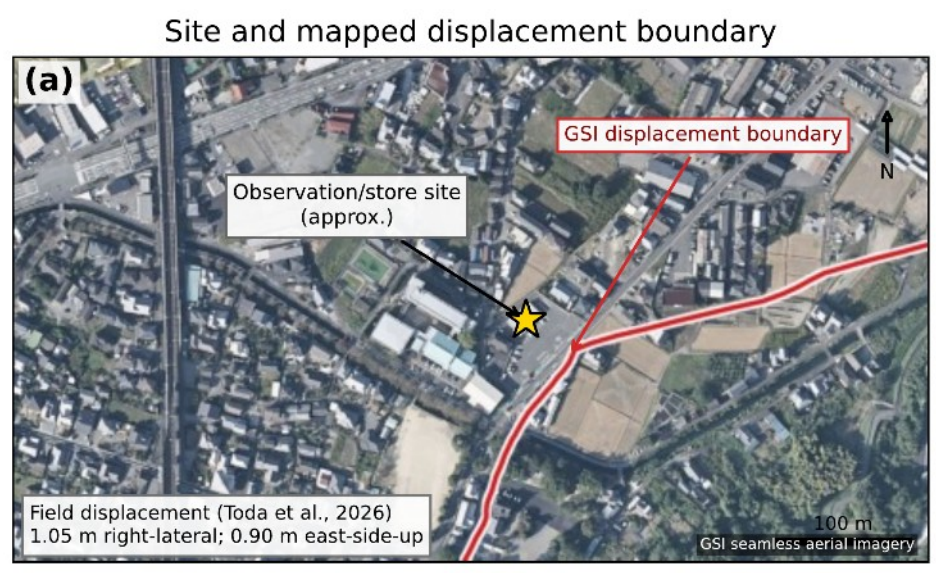

Field expression of the surface rupture

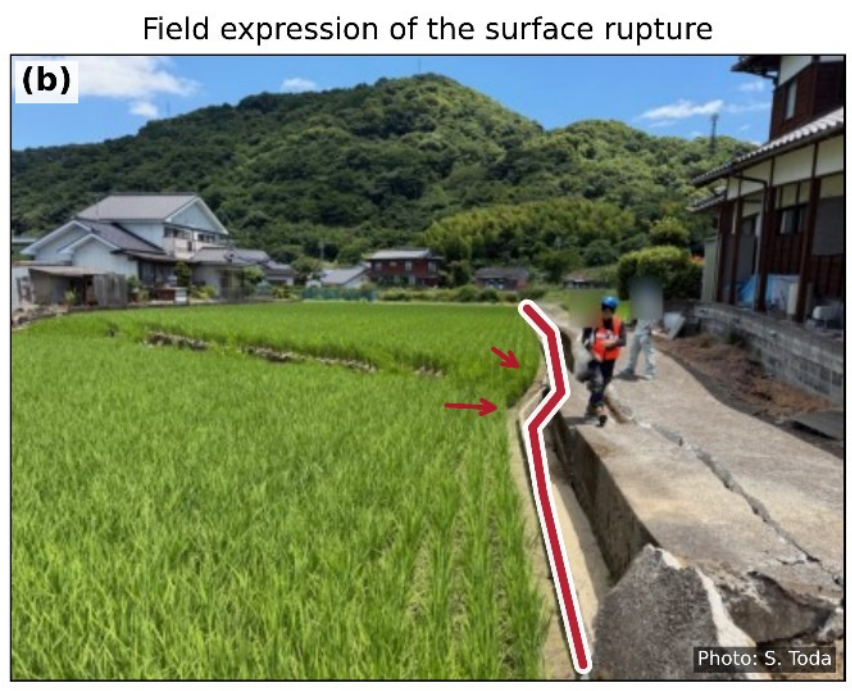

Pre-slip frame: near-side references and far-side target

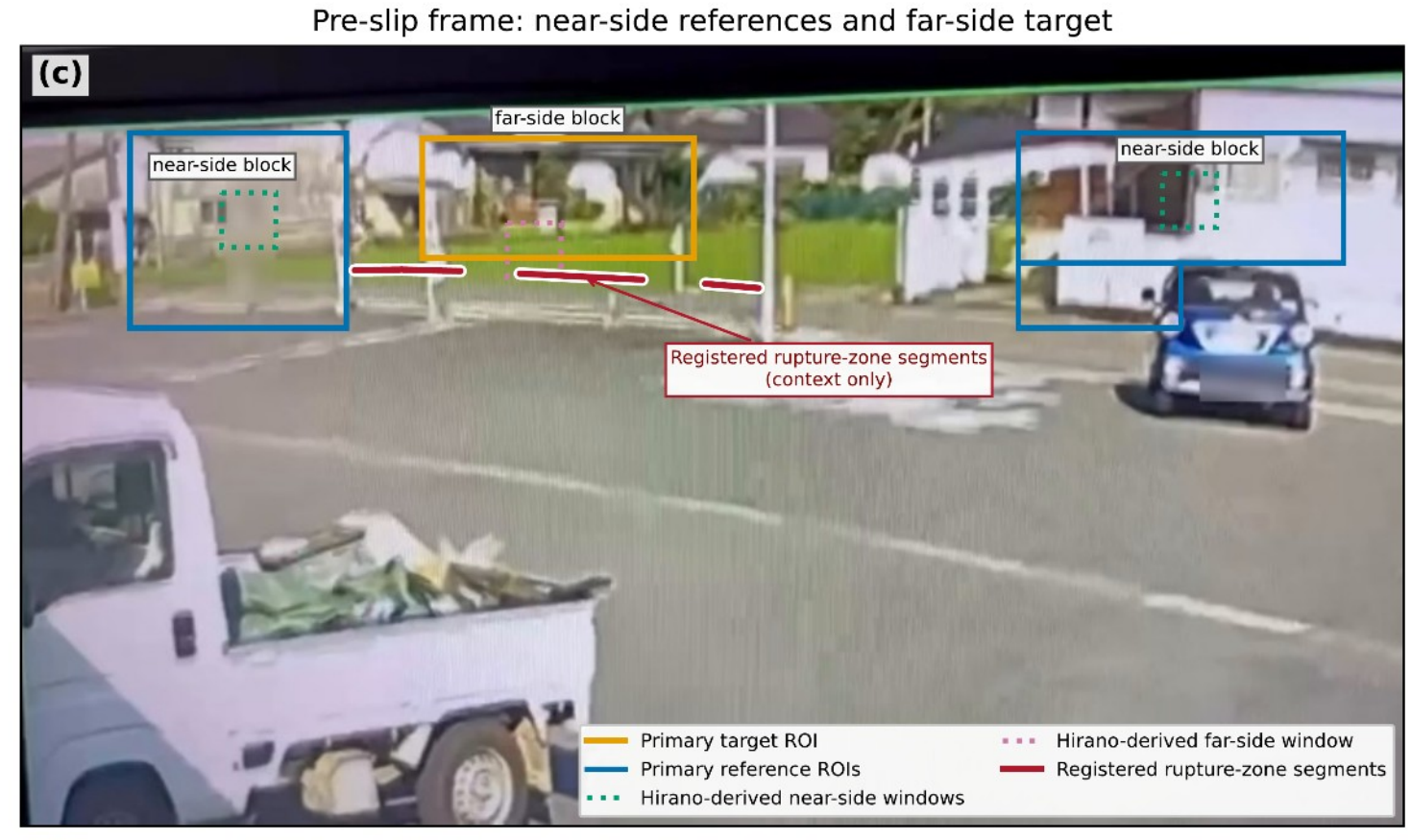

**Figure 3**

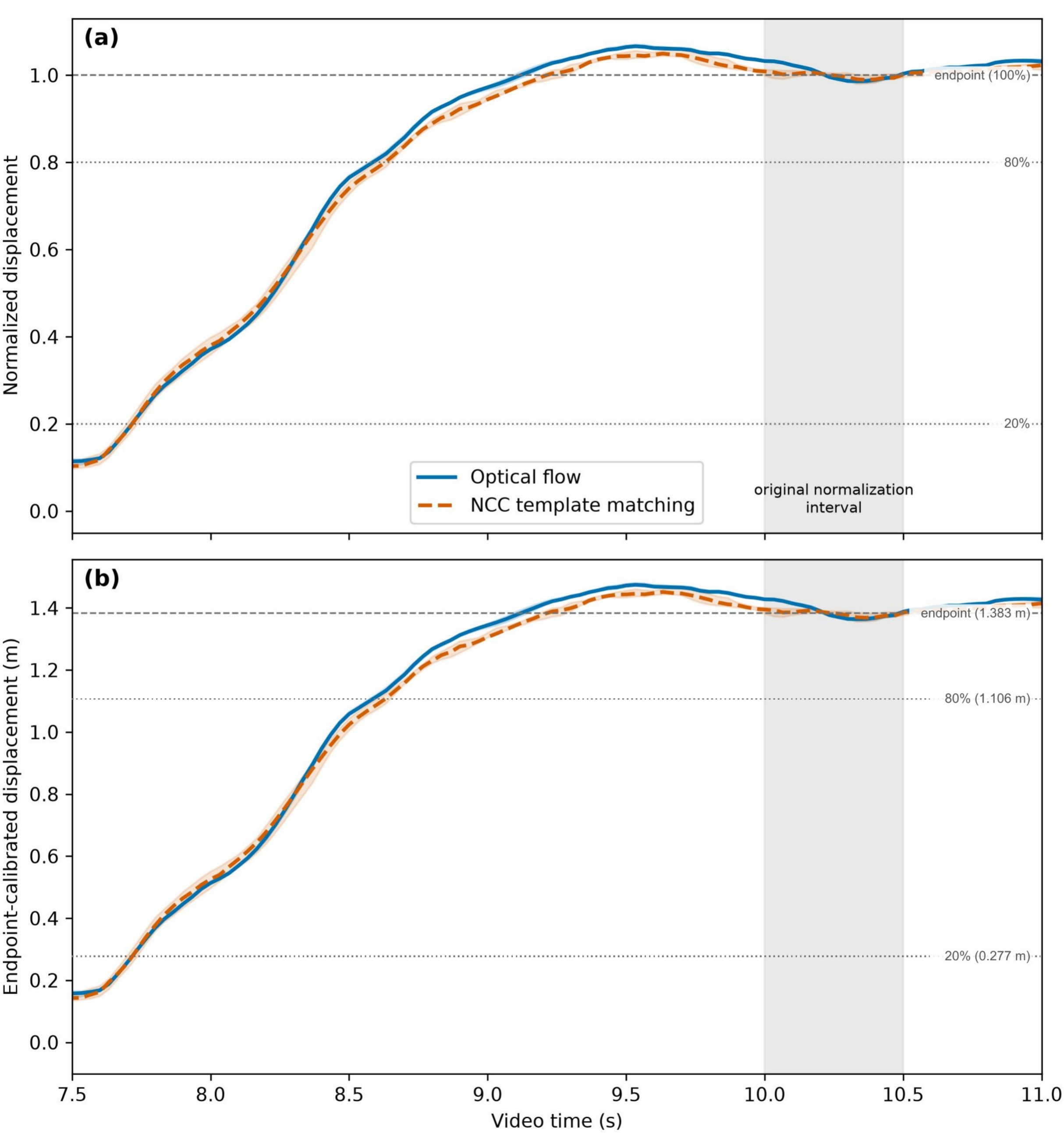
(a)
1.0
0.8
0.6
0.4
0.2
0.0
Normalized displacement
endpoint (100%)
80%
20%
Optical flow
NCC template matching
original normalization interval
(b)
1.4
1.2
1.0
0.8
0.6
0.4
0.2
0.0
Endpoint-calibrated displacement (m)
endpoint (1.383 m)
80% (1.106 m)
20% (0.277 m)
7.5
8.0
8.5
9.0
9.5
10.0
10.5
11.0
Video time (s)

## Figure 4

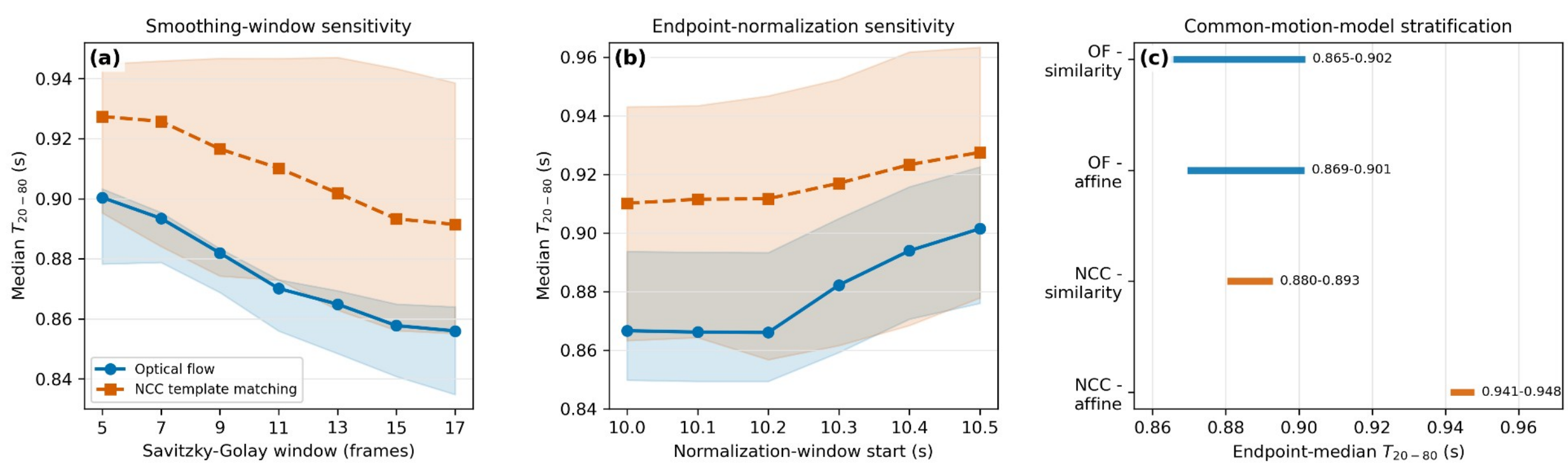

Smoothing-window sensitivity
(a)
Median $T_{20-80}$ (s)
0.94
0.92
0.90
0.88
0.86
0.84
Optical flow
NCC template matching
5
7
9
11
13
15
17
Savitzky-Golay window (frames)
Endpoint-normalization sensitivity
(b)
Median $T_{20-80}$ (s)
0.96
0.94
0.92
0.90
0.88
0.86
0.84
10.0
10.1
10.2
10.3
10.4
10.5
Normalization-window start (s)
Common-motion-model stratification
(c)
OF - similarity
OF - affine
NCC - similarity
NCC - affine
0.865-0.902
0.869-0.901
0.880-0.893
0.941-0.948
0.86
0.88
0.90
0.92
0.94
0.96
Endpoint-median $T_{20-80}$ (s)

**Figure 5**

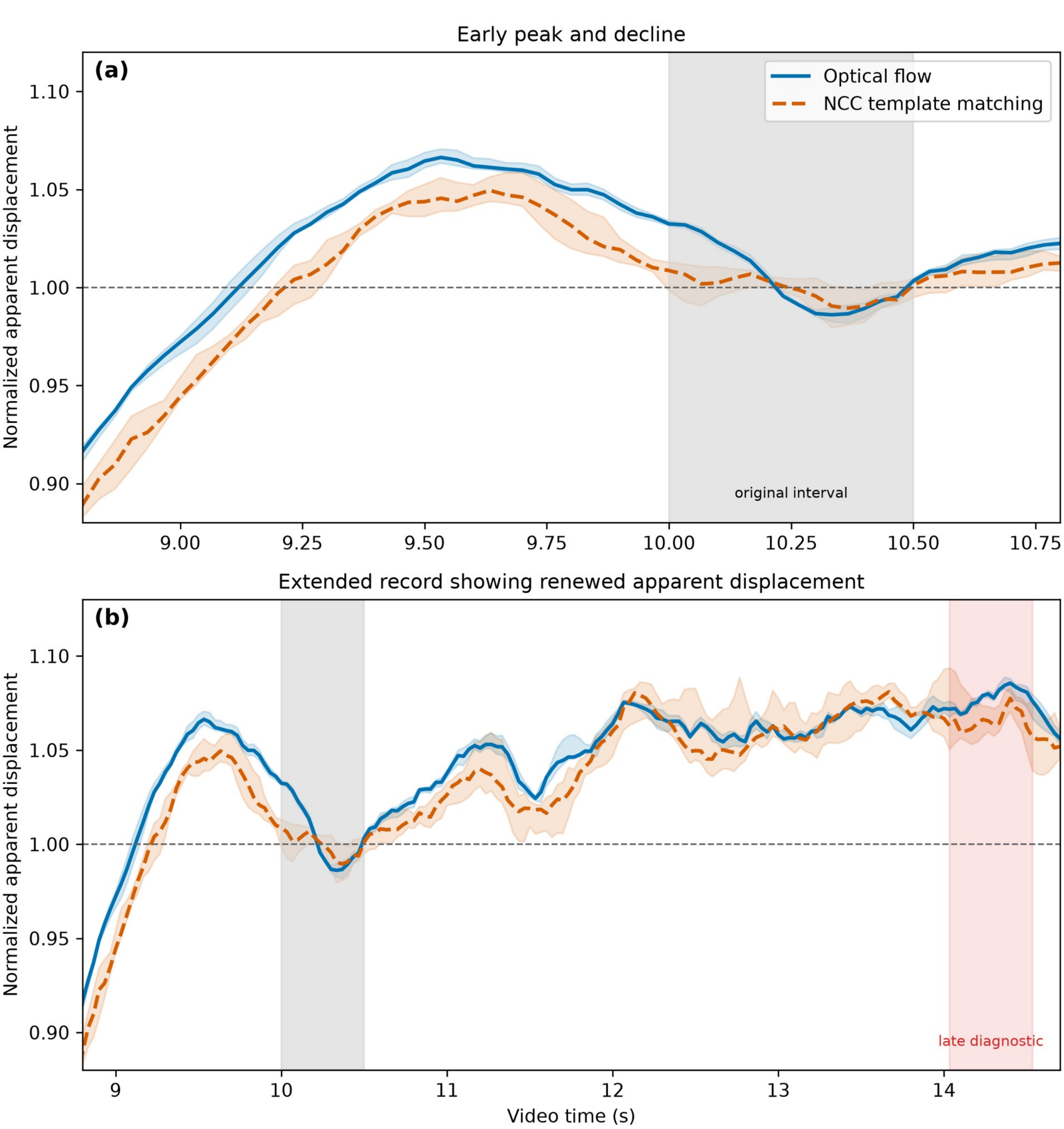

Early peak and decline
(a)
Optical flow
NCC template matching
original interval
Normalized apparent displacement
1.10
1.05
1.00
0.95
0.90
9.00
9.25
9.50
9.75
10.00
10.25
10.50
10.75
Extended record showing renewed apparent displacement
(b)
late diagnostic
Normalized apparent displacement
9
10
11
12
13
14
Video time (s)

# Supporting Information

## Time-Resolved Surface-Fault Displacement During the 2026 Kumamoto Earthquake From Near-Fault Video

This Supporting Information contains Texts S1–S15, Tables S1–S10, and Figures S1–S12.

The main paper presents the seismological result. This supplement records the exact image-processing settings, sensitivity tests, and unsuccessful or inconclusive diagnostics needed to evaluate that result. Figure S12 gives a plain-language overview of the complete workflow; Texts S3–S5 document tracking and calibration; and Texts S6–S15 report robustness checks and limitations.

# Contents

## Text S1. Video characteristics and coordinate convention

The analyzed video is a secondary smartphone recording of a closed-circuit television (CCTV) monitor, not the native security-camera file. According to information obtained from the video provider through Yomiuri Shimbun, the store system automatically deleted the native recording after ten days, and it could not be recovered. The copy used for every analysis contains 442 frames at a nominal rate of 30 frames $s^{-1}$, has a spatial resolution of 1276 × 720 pixels, and lasts 14.733 s. The image is four pixels narrower than the nominal 1280 × 720 format; the cause cannot be determined from the secondary copy. Spatially resized copies were not used for measurement.

The exact video file used for all final analyses has a SHA-256 checksum of F3DD985B4D31EF5DBE77822C46E084A088149E5AF098557F2954676FDF73FB51. Its encoded H.264 stream contains 442 frames with uniformly spaced presentation timestamps at a nominal interval of 1/30 s. The 5.0–11.0-s analysis interval therefore contains 181 frames. The stream was decoded without errors, and no exact duplicate adjacent frames were found. These checks support the use of the encoded 1/30-s frame spacing, but they do not verify the timing of the native security-camera recording or the monitor and smartphone capture. Timing errors caused by display refresh, rolling shutter, or transcoding cannot be quantified from the secondary file alone and would be shared by both tracking methods.

The predefined primary analysis uses video times 5.0–11.0 s (181 frames). The endpoint diagnostic extends through the final usable frames, with a late comparison interval of 14.033–14.533 s chosen before examining the diagnostic output and ending five frames before the final frame. Pixel coordinates increase rightward and downward; upward displacement is shown positive.

The primary target ROI was fixed at x = 370–620 and y = 85–195 pixels. The reference regions were: left, x = 100–300 and y = 80–260; right upper, x = 920–1220 and y = 80–200; and right lower, x = 920–1070 and y = 200–260. The lower-right parked vehicle was excluded. Representative target features and templates are shown in Figure S11; most lie on the distant structures because the paddy surface has little stable texture.

## Text S2. Registration of the rupture-zone segments shown in Figure 2c

The source image is the lower post-motion Yomiuri still retained in the reproducibility archive as Fault_regionX_3.jpg. Eight points were picked manually along the visible dark surface break within the red published annotation. These points define the displayed rupture-zone segments; they are not surveyed coordinates or registration controls.

Registration used an automated scale-invariant feature transform (SIFT) and a projective homography. The source-image mask excluded the red annotation, image border, and bottom caption. Detection yielded 1,148 source and 2,317 target keypoints; Lowe-ratio filtering at 0.72 retained 42 candidate matches. Random sample consensus (RANSAC) fitting used a 4.0-pixel reprojection threshold, 10,000 maximum iterations, and 0.999 confidence, retaining 19 geometrically consistent registration inliers. These inliers were selected algorithmically rather than as manual stationary controls. Their reprojection residuals have a 1.37-pixel root-mean-square value, 1.01-pixel median, and 3.02-pixel maximum.

The 19 inliers span x = 306–568 and y = 102–240 pixels in the pre-slip frame. Seventeen lie within the primary far-side target ROI; the remaining two lie outside both the target and reference ROIs and were not assigned to either block. The homography is therefore predominantly far-side anchored. It supports local image registration around the displayed far-side segment but does not independently constrain the near-side reference regions or establish the near-side/far-side block assignment.

The 1.37-pixel residual measures control-point agreement only and is not a complete uncertainty on the geological trace. Neither the original camera pose nor the full three-dimensional scene geometry was surveyed. Outside the spatial support of the inliers, the transfer is unconstrained, so the segments are terminated rather than extrapolated toward the reference regions. They provide display-only geometrical context and are not inputs to feature selection, common-motion fitting, or any reported displacement quantity.

## Text S3. Optical-flow processing ensemble

Optical flow estimates image displacement by following identifiable features between frames. The final analysis used Shi–Tomasi features and pyramidal Lucas–Kanade tracking directly between the baseline frame at video time 5.0 s and each subsequent frame. Direct baseline-to-frame tracking was used rather than sequential frame-to-frame accumulation to reduce long-term drift.

Lucas–Kanade parameters were a 31 × 31 pixel tracking window, maximum pyramid level of 5, maximum of 50 iterations, termination tolerance of 0.01 pixels, and a forward–backward error threshold of 1.5 pixels.

Shi–Tomasi feature selection used a quality threshold of 0.01, a minimum spacing of 7 pixels, and a block size of 5 pixels. Up to 1000 features were selected for the target region and for the feature-weighted reference set. For the patch-balanced reference scheme, the same selection thresholds were applied independently within each of the 12 fixed tiles, with up to 100 features per tile.

Two reference-motion weighting schemes were considered. In the feature-weighted scheme, all accepted reference features contributed directly to estimation of the common transformation. In the patch-balanced scheme, the reference area was divided into 12 fixed spatial tiles, six on each side of the target region, and each tile contributed a representative displacement. This reduced the influence of unequal feature density among different parts of the image.

For each weighting scheme, common image motion was represented by either a similarity transformation (translation, rotation, and uniform scaling) or an affine transformation (also allowing nonuniform scaling and shear), giving four optical-flow variants (Table S1). For each frame, the fitted common-motion transformation predicted where the baseline target features would appear in the absence of target-specific displacement. The robust median residual between observed and predicted target positions defined the target displacement. Common-motion fitting used RANSAC with a 2.5-pixel residual threshold, 5,000 maximum iterations, 0.999 confidence, and 10 refinement iterations.

## Text S4. Normalized cross-correlation (NCC) template-matching processing ensemble

Normalized cross-correlation (NCC) provided an independent tracking method. Template centers were selected from the baseline frame using Shi–Tomasi features only to identify well-textured image regions. For the reference regions, up to 36 centers were selected using a quality threshold of 0.02, minimum spacing of 32 pixels, and block size of 5 pixels. For the target region, up to 18 centers were selected using the same quality threshold and block size and a minimum spacing of 28 pixels.

Templates were tracked using OpenCV TM_CCOEFF_NORMED within a search radius of 90 pixels. Matches with normalized correlation coefficients below 0.55 were rejected. The correlation maximum was refined to subpixel precision by quadratic interpolation.

Three template sizes were tested: 31 × 31, 41 × 41, and 51 × 51 pixels. For each template size, common recording motion was modeled using either a similarity or affine transformation, giving six fixed final NCC variants (Table S2). Common-motion fitting used the same frozen RANSAC settings as optical flow: a 2.5-pixel residual threshold, 5,000 maximum iterations, 0.999 confidence, and 10 refinement iterations. The residual displacement at the target-template positions after removal of the fitted common motion was summarized using the median. At the original 10.0–10.5-s interval, the two central $T_{20–80}$ values of the pooled 42-realization ensemble are 0.909275 s (41-pixel similarity, 11-frame smoothing) and 0.911047 s (51-pixel affine, 11-frame smoothing), giving a median of 0.910161 s.

## Text S5. Common postprocessing and image-to-ground calibration

Optical-flow and NCC histories were processed in the same way. A second-order Savitzky–Golay filter used seven fixed odd windows: 5, 7, 9, 11, 13, 15, and 17 frames, corresponding to 0.17–0.57 s at the nominal 30 frames $s^{-1}$. The pre-event baseline was 5.0–6.5 s. To test sensitivity to the final reference level, six normalization intervals were used from 10.0–10.5 through 10.5–11.0 s in 0.1-s steps. The median displacement in each interval defined 100% for that realization; no interval is treated as a demonstrated settled plateau.

For every realization and endpoint interval, the baseline was removed, the interval-defined two-dimensional endpoint vector was recomputed, the history was projected onto that direction, and the scalar displacement was normalized by its interval median. Crossing times were then recomputed by linear interpolation after 6.5 s; the test was not a cosmetic rescaling of a fixed curve.

$$T_{20–80} = t_{80} - t_{20}$$

The 20–80% duration was used as the primary measure of the principal displacement timescale, whereas the 10–90% duration was retained as a secondary measure because the low-amplitude beginning and end of the displacement are more sensitive to residual image motion and smoothing.

Scalar calibration used the independently measured permanent-displacement components of 1.05 m right-lateral and 0.90 m east-side-up, giving a field-vector magnitude $D_f = 1.38293$ m. For each processing realization, one empirical scale factor was obtained by matching the final image-vector magnitude $|D_{px}|$ to $D_f$:

$$s = D_f / |D_{px}|$$

where $|D_{px}|$ is the interval-defined image-vector magnitude. The same scale factor is held fixed throughout that realization. This is an empirical calibration rather than a photogrammetric reconstruction, and it assumes that the effective image scale remains sufficiently constant during the main rise. Movement of the phone relative to the monitor could violate this assumption. The average rate over the central 20–80% interval is

$$V_{20\text{–}80} = 0.60\ D_f / T_{20\text{–}80} = 0.829759\ \text{m} / T_{20\text{–}80}.$$

The image endpoint vector cancels exactly from this scalar average rate. Endpoint choice can still alter the projected direction and crossing times, and a time-varying geometric distortion could change the shape of the history. Processing and normalization spreads do not include biases shared by all analyses of the secondary recording. The field team estimates an approximate ±0.05 m tape-measurement uncertainty for each measured component. Treating the component uncertainties as independent gives approximately ±0.050 m (±3.6%) on $D_f$ and on the calibrated rate. This calibration uncertainty is reported separately. The image-component ratios and their agreement with the field-vector direction are discussed in the main text and Text S14.

## Text S6. Sensitivity of displacement timing to processing choices

Across the six original endpoint intervals, method-median $T_{20\text{–}80}$ values range from 0.866 to 0.901 s for optical flow and from 0.910 to 0.928 s for NCC (Table S7; Figure S6). At 10.0–10.5 s, the values are 0.867 [0.850–0.894] s and 0.910 [0.863–0.943] s. Across the full 5–17-frame smoothing range, the method median changes by 0.044 s for optical flow and 0.036 s for NCC, approximately 5% and 4% of the main duration. Smoothing therefore affects the exact timing but does not generate the rapid rise. Target-ROI perturbations produce smaller median paired changes of 0.005–0.008 s (Text S11).

Main Figure 4a shows the smoothing sensitivity of the primary 20–80% duration. Figure S2 shows the secondary 10–90% duration, and Figure S3 shows the stronger smoothing dependence of the instantaneous peak rate. In the primary geometry, common-motion model choice has little effect on optical flow: similarity gives 0.865–0.902 s and affine 0.869–0.901 s, with paired affine-minus-similarity differences from −0.003 to +0.005 s (median +0.002 s). NCC separates by model: similarity gives 0.880–0.893 s and affine 0.941–0.948 s, with paired differences of +0.055 to +0.064 s (median +0.060 s; Table S10). We therefore keep $T_{20\text{–}80}$ as the primary timing measure and report the NCC model groups separately rather than treating the pooled median as an individual realization.

## Text S7. Sensitivity of instantaneous peak displacement rate

Instantaneous displacement rate was calculated from the first derivative of the same second-order Savitzky–Golay representation used for the displacement histories. In contrast to the 20–80% average rate, the peak-rate amplitude varies appreciably with smoothing (Table S4; Figure S3).

For optical flow, median peak-rate estimates decrease from 1.591 m s$^{-1}$ at the 5-frame window to 1.205 m s$^{-1}$ at the 17-frame window. NCC shows similar behavior and substantially larger variation at the shortest smoothing window, where the 16th–84th percentile range is 1.501–2.190 m s$^{-1}$.

Across the complete fixed final processing spread, the optical-flow peak rate is 1.428 [1.271–1.556] m s$^{-1}$ and the NCC peak rate is 1.275 [1.135–1.482] m s$^{-1}$. These results show that the existence of rapid motion is robust, whereas the amplitude of the instantaneous maximum is smoothing-sensitive. Accordingly, the mean 20–80% rate is retained as the primary rate measure in the main text.

## Text S8. Spatial and reference-region control tests

The reference geometry was divided into six fixed horizontal zones: L1, x = 100–300 and y = 80–140 pixels; L2, x = 100–300 and y = 140–200; L3, x = 100–300 and y = 200–260; R1, x = 920–1220 and y = 80–140; R2, x = 920–1220 and y = 140–200; and R3, x = 920–1070 and y = 200–260. Each zone comprises two of the 12 patch-balanced reference tiles. Some nominal stationary controls, particularly on the right side, contained too little well-conditioned structure to provide a stable independent time history.

The unstable controls could produce excursions larger than the target's local early excess while also retaining large final residuals, so they could not bound recording artifacts quantitatively. A display-border control was also non-identifying: baseline support was adequate, but the border had insufficient horizontal texture and late tracking failed.

These controls are retained as failed or non-identifying diagnostics. They establish neither the physicality nor the absence of the late apparent motion.

## Text S9. Reference-region jackknife analysis

To test whether the early peak and decline depended on one part of the reference image, the full analysis was repeated after omitting each of L1–L3 and R1–R3 while holding the target fixed.

The local early excess relative to the original normalization interval remains positive for every omission in both methods (Figure S4). R1 produces the largest spread, but no single zone alone creates the peak/decline pattern.

This is an internal robustness result for the processed video history. It cannot identify physical overshoot because every realization uses the same secondary recording and the extended histories later rise above the earlier peak.

## Text S10. Within-video early peak-and-decline classification

The frozen classification searches for a local maximum from each realization's $t_{80}$ through 10.0 s and evaluates subsequent negative velocity before the original 10.0–10.5-s interval. It therefore tests whether an early peak and decline are resolved within that part of the video, not whether a permanent plateau or physical overshoot exists.

For full-reference processing, the local excess above the original interval is 6.626% for optical flow and 5.461% for NCC, with maxima near 9.6 s. Text S12 gives values up to 0.04 percentage point different because its pipeline removes the baseline and aggregates components in a different order. Both calculations use the same frozen input data and settings. Table S5 and Figure S5 retain the peak/decline-classification values.

A realization satisfies the internal peak/decline criterion when the local excess exceeds three times pre-event displacement noise and at least N consecutive velocity samples exceed K times pre-event velocity noise with positive signs before and negative signs after the local maximum. K = 0, 1, or 2 and N = 2, 3, or 4 were frozen before evaluation.

For K = 1 and N = 3, the classification is met by 96.4% of full-reference optical-flow cases, 100% of full-reference NCC cases, 95.8% of optical-flow jackknife cases, and 99.2% of NCC jackknife cases. Stricter criteria reduce these fractions (Figure S5).

The classification shows that an early peak and decline are robust features of the video-derived history under the original normalization. It is deliberately not called a physical overshoot test: the late level lies above both the original interval and the early peak, and coherent secondary-recording geometry remains possible.

## Text S11. Sensitivity to target-region definition

To evaluate whether the NCC timing depends on the analyst-defined target region, we performed two diagnostic perturbation experiments after the fixed final NCC ensemble had been established. In both experiments, the reference regions and all NCC tracking and postprocessing parameters were unchanged. For each template size, common-motion model, and frame, the reference matches and common-motion transformation were calculated once and then frozen across the target-region perturbations. Target Shi–Tomasi template centers were redetected independently within each perturbed ROI. Each target ROI was evaluated using all three template sizes (31, 41, and 51 pixels), both common-motion models (similarity and affine), and all seven smoothing windows, giving 42 paired processing realizations per ROI.

For the translation test, the median absolute paired change in $T_{20–80}$ is 0.005 s (84th percentile 0.015 s; maximum 0.064 s; Table S6). ROI-level medians range from 0.887 to 0.913 s. The local early-excess percentage changes by a median 0.282 percentage points but is not treated as a physical endpoint quantity.

For the size test, the median absolute paired change in $T_{20–80}$ is 0.008 s (84th percentile 0.019 s; maximum 0.050 s; Table S6). ROI-level medians range from 0.897 to 0.914 s. The local early-excess percentage changes by a median 0.409 percentage points.

These tests support the approximately 0.9-s principal timescale and show that it is less sensitive to target-ROI definition than the few-percent early-excess quantity.

## Text S12. Endpoint-normalization and late-extension test

Six 0.5-s original endpoint intervals beginning from 10.0 through 10.5 s were defined before the test. Each interval recomputed endpoint direction, normalization, and crossings. Optical-flow median $T_{20–80}$ increases from 0.867 to 0.901 s and NCC from 0.910 to 0.928 s across the grid. The local early excess decreases from 6.642–6.667% to 4.640% for optical flow and from 5.468–5.491% to 4.241% for NCC (Figure S6). The corresponding factors, 1.44 and 1.29, do not exceed the prespecified 1.5× threshold for strong dependence within the primary grid; the sign reversal appears only when the exploratory higher late level is used.

The late interval 14.033–14.533 s is exploratory. It produces $T_{20–80}$ = 0.972 s (optical flow) and 0.998 s (NCC). In the method-median 11-frame histories, the late level lies 7.880% and 6.325% above the original 10.0–10.5-s level and 1.196% and 1.466% above the earlier apparent peak for optical flow and NCC, respectively. These method-median-curve values are retained to describe the plotted histories; the record does not demonstrate settlement by 10.0–10.5 s.

For a same-aggregation comparison of normalization definitions, each tracker realization was smoothed with the fixed 11-frame window, projected along its own endpoint direction, normalized to the stated interval, and evaluated before taking the median across tracker variants. The earlier apparent peak was searched from each realization's $t_{80}$ crossing through 10.0 s, and the late level was the median over 14.033–14.533 s. With the original endpoint direction and 10.0–10.5-s normalization, the late level exceeds the earlier peak by 0.999% for optical flow [16th–84th percentile across realizations, 0.549–1.434%] and 1.149% for NCC [−0.088–1.895%]. With the late endpoint direction and 14.033–14.533-s normalization, the values are 1.045% [0.536–1.475%] and 1.112% [0.194–1.774%], respectively.

These same-aggregation values must not be obtained by subtracting rounded headline summaries, because the operation order differs. The full-precision calculation does not show the anticipated method-order reversal: NCC is slightly larger than optical flow under both definitions. The apparent reversal arose from comparing summaries formed with different aggregation orders. The late rise above the earlier apparent peak is qualitatively common to both method histories, but its approximately 1% magnitude is sensitive to projection, normalization, and aggregation.

## Text S13. Geometry-restricted validation

Hirano supplied the exact inclusive coordinates of his two fixed 50 × 50-pixel windows—left (185, 135)–(234, 184) and right (1053, 117)–(1102, 166)—and tracked window (448, 163)–(497, 212). Coordinates are relative to the upper-left pixel of the 90th frame of his 180-frame segment beginning at video time 5.0 s. We froze these coordinates before rerunning the geometry check (Figure S7). All pixels outside the three windows were excluded from this test.

Optical flow accepted all 29 near-side reference candidates and all 16 far-side target candidates (Table S8). All 181 frames were valid. Surviving reference counts ranged from 8 to 29 (median 24), and RANSAC inliers ranged from 6 to 29, with medians of 19 for affine and 17 for similarity correction. Across the seven smoothing windows, similarity $T_{20–80}$ is 0.826–0.852 s and affine $T_{20–80}$ is 0.842–0.892 s. The ranges overlap, and at 11-frame smoothing the values are 0.842 and 0.850 s (Figure S8). Thus, the factor-of-two split found with the earlier approximately digitized windows is not reproduced.

The unchanged multi-template NCC pipeline remained non-viable inside the exact small windows. At 31 and 41 pixels, only two reference templates and one target template were fully contained; 51 pixels retained none. This was insufficient for the documented common-motion fit, so all 42 restricted NCC variants remain invalid. Frame-by-frame counts and inliers are preserved in the reproducibility archive.

## Text S14. Corrected whole-window NCC cross-check

An earlier whole-window NCC run used crops from the 5.0-s baseline frame. Hirano subsequently clarified that his templates came from the 90th frame of the 180-frame segment, corresponding to video frame 239 at 7.967 s. The baseline-crop run is preserved as a failed provenance diagnostic and is not interpreted.

The corrected whole-window NCC cross-check uses Hirano's exact coordinates and central-frame templates. All 181 frames pass the unchanged quality rule. The endpoint is 24.868 px, within 0.11% of the exact-window optical-flow endpoint (24.841 px) and 0.10% of the primary NCC endpoint (24.844 px). $T_{20–80}$ is 0.826 s for manuscript-compatible 11-frame smoothing and 0.846 s for Hirano-compatible weighting and smoothing (Figure S9). These values are shorter than primary NCC and are reported as a same-recording cross-check, not merged with the primary ranges.

The apparent early excess is 6.423% for the 11-frame history and 6.326% for the Hirano-compatible history. Their peak significances are 2.281σ and 2.673σ, respectively, so both fail the frozen >3σ gate. This does not weaken the meter-scale main displacement; it means the single-window cross-check does not independently establish the small early-excess feature.

Directions were calculated for each realization before aggregation. Primary optical flow is 40.749° [40.405–40.906°], primary NCC is 40.391° [39.997–40.621°], and the field vector is 40.601°. The exact-window optical-flow direction is 36.662° [36.275–37.357°], while whole-window NCC gives 47.656°. Local windows therefore reproduce endpoint magnitude more closely than direction, because each samples one structure and image depth.

These window-based checks use the same secondary recording and are not independent displacement observations. Their value is to test geometry and processing behavior, not to replace the spatially distributed primary ensemble.

## Text S15. Projective and display-border artifact diagnostics

A reference-only homography retained 1.274 px (58.0%) of the 2.197-px optical-flow late increment, with coherent target strips, but retained −0.079 px (−3.8%) of the 2.076-px NCC increment, with incoherent strips. Held-out reference validation failed for both methods (Figure S10). The result is therefore mixed and does not validate a correction for the earthquake history.

A screen-border control passed its baseline-support check but lacked horizontal texture and stable late tracking. It therefore cannot distinguish earthquake motion from phone-to-display motion.

We next tested the two-stage recording geometry suggested by Nobuhara: estimate phone motion only from the flat monitor hardware, without using the earthquake scene. Controlled tests showed why this method could not yet be applied reliably. Direct homography alignment in normalized coordinates recovered broad synthetic texture accurately, but the two real bezel views had median maximum-grid errors of 0.864 and 4.669 px. A four-corner displacement parameterization performed worse on the same views (1.237 and 6.622 px, with 84th-percentile errors near 14 px). The visible bezel therefore provides weak or uneven information. These tests do not validate a correction for the movie.

An exploratory row-wise stabilization supplied by Nobuhara uses the green display line as its only hard constraint. It fixes the line's position and slope while estimating the other six homography degrees of freedom under a soft image prior. The prior is proportional to $(\text{inlier rate})^3 \times (1 + 6 \times \text{ROI})$, where ROI is a 150-pixel band below the green line together with bright pixels on the road surface; it does not encode near-side or far-side membership. Because this region overlaps 57.9% of the primary far-side target ROI, the stabilization could remove part of the earthquake signal. The updated 720p result visibly reduces large phone-to-monitor motion, but preservation of the fault displacement has not been demonstrated. We therefore use it only as a diagnostic, and no reported displacement was calculated from the stabilized video. None of the projective tests provides a numerical bound on frame-timing errors or time-varying geometric distortion shared by the entire secondary recording. The late apparent motion remains unresolved, whereas the primary approximately 0.9-s rise remains unchanged. Figure S12 separates the primary measurement from these artifact diagnostics and summarizes the workflow.

References cited in this Supporting Information are included in the main paper's reference list.

## Supporting Tables

**Table S1.** Optical-flow processing variants

| Variant | Reference weighting | Common-motion model |
|---|---|---|
| OF1 | Feature-weighted | Similarity |
| OF2 | Feature-weighted | Affine |
| OF3 | Patch-balanced | Similarity |
| OF4 | Patch-balanced | Affine |

**Note.** *All optical-flow variants used Shi–Tomasi feature selection and pyramidal Lucas–Kanade tracking. Common recording motion was estimated from the reference regions and removed from the target-region motion.*

**Table S2.** NCC template-matching processing variants

| Variant | Template size | Common-motion model |
|---|---|---|
| NCC1 | 31 px | Similarity |
| NCC2 | 31 px | Affine |
| NCC3 | 41 px | Similarity |
| NCC4 | 41 px | Affine |
| NCC5 | 51 px | Similarity |
| NCC6 | 51 px | Affine |

**Note.** *All NCC variants used normalized cross-correlation tracking of fixed templates selected from the baseline frame. Common recording motion was estimated independently from the reference regions and removed from the target-region motion.*

**Table S3.** Original 10.0–10.5-s processing spread of principal and descriptive quantities

| Quantity | Optical flow | NCC template matching |
|---|---|---|
| Final vector (px) | 24.280 [23.996–24.389] | 24.844 [24.758–24.948] |
| Scalar endpoint scale (cm $px^{-1}$) | 5.696 [5.670–5.763] | 5.566 [5.543–5.586] |
| 20–80% rise (s) | 0.867 [0.850–0.894] | 0.910 [0.863–0.943] |
| 10–90% rise (s) | 1.381 [1.324–1.443] | 1.517 [1.377–1.650] |
| $t_{50}$ (s) | 8.224 [8.218–8.236] | 8.210 [8.190–8.250] |
| Mean 20–80% rate (m $s^{-1}$) | 0.957 [0.928–0.976] | 0.912 [0.880–0.961] |

| Quantity | Optical flow | NCC template matching |
|---|---|---|
| Local early excess above original interval (%) | 6.626 [6.212–7.282] | 5.461 [4.996–7.008] |
| Local early-peak time (s) | 9.567 [9.533–9.600] | 9.600 [9.519–9.733] |

Note. Values are medians [16th–84th percentiles] at the original 10.0–10.5-s interval. They are processing spreads, not statistical confidence intervals. The last two rows describe the early video-history peak relative to that normalization choice and are not evidence of physical overshoot. The local early-excess and peak-time rows were produced by the peak/decline-classification pipeline described in Text S10, which projects and baseline-centers the scalar history before smoothing.

**Table S4.** Sensitivity of peak displacement rate to smoothing

**(a) Peak rate by smoothing window**

| Smoothing window | Optical flow peak rate (m $s^{-1}$) | NCC peak rate (m $s^{-1}$) |
|---|---|---|
| 5 frames | 1.591 [1.538–1.666] | 1.679 [1.501–2.190] |
| 7 frames | 1.553 [1.496–1.565] | 1.398 [1.345–1.516] |
| 9 frames | 1.440 [1.432–1.455] | 1.313 [1.297–1.369] |
| 11 frames | 1.430 [1.414–1.433] | 1.251 [1.218–1.290] |
| 13 frames | 1.365 [1.350–1.382] | 1.219 [1.166–1.269] |
| 15 frames | 1.279 [1.272–1.295] | 1.165 [1.093–1.212] |
| 17 frames | 1.205 [1.196–1.217] | 1.098 [1.033–1.162] |

**(b) Complete fixed final processing spread**

| Quantity | Optical flow | NCC template matching |
|---|---|---|
| Peak rate (m $s^{-1}$) | 1.428 [1.271–1.556] | 1.275 [1.135–1.482] |
| Full range (m $s^{-1}$) | 1.192–1.712 | 0.961–2.469 |
| Peak time (s) | 8.333 [8.300–8.367] | 8.333 [7.733–8.367] |

Note. Peak-rate amplitude is substantially more sensitive to smoothing than the mean 20–80% rate reported in Table S3 and is therefore treated as a secondary result.

**Table S5.** Within-video early peak-and-decline classification

(a) Local early peak-and-decline summary

| Quantity | Optical flow (full reference) | NCC (full reference) | Optical flow (reference jackknife) | NCC (reference jackknife) |
|---|---|---|---|---|
| Peak time (s) | 9.567 [9.533–9.600] | 9.600 [9.519–9.733] | 9.567 [9.500–9.667] | 9.567 [9.467–9.667] |
| Local early excess (%) | 6.626 [6.212–7.282] | 5.461 [4.996–7.008] | 6.828 [6.053–9.374] | 6.014 [5.291–8.173] |

| Quantity | Optical flow (full reference) | NCC (full reference) | Optical flow (reference jackknife) | NCC (reference jackknife) |
|---|---|---|---|---|
| Peak-to-10.0 s decrease (%) | 3.324 [2.910–3.838] | 4.594 [3.893–6.713] | 3.467 [2.722–5.487] | 4.846 [3.741–6.932] |
| Peak excess / baseline noise (σ) | 9.315 [7.491–11.147] | 5.776 [4.642–9.052] | 8.894 [6.395–13.129] | 8.167 [5.085–13.174] |

(b) Fraction satisfying the internal peak-and-decline criterion

| Criterion | Optical flow (full) | NCC (full) | Optical flow (jackknife) | NCC (jackknife) |
|---|---|---|---|---|
| $K = 1\sigma_v$, $N = 3$ | 96.4% | 100.0% | 95.8% | 99.2% |
| $K = 2\sigma_v$, $N = 3$ | 78.6% | 95.2% | 73.8% | 84.9% |

Note. This criterion measures an early peak and subsequent decline within the original normalized video history. It does not require or establish a permanent plateau and is not a physical-overshoot test. Values in both parts were produced by the peak/decline-classification pipeline described in Text S10, including scalar projection and baseline centering before smoothing.

**Table S6.** NCC target-region definition sensitivity

| Test | Perturbation | **Paired** $\|\Delta T_{20–80}\|$: median / p84 / max (s) | **ROI-level median** $T_{20–80}$ range (s) |
|---|---|---|---|
| Translation | 25 ROIs; ±10 px in 5-px steps in x and y | 0.005 / 0.015 / 0.064 | 0.887–0.913 |
| Size | 5 centered ROIs; 80–120% of baseline dimensions | 0.008 / 0.019 / 0.050 | 0.897–0.914 |

Note. Paired differences compare each perturbed ROI with the baseline ROI for identical template size, common-motion model, and smoothing. Reference transformations were frozen. The tests are diagnostic and are not included in the primary normalization-grid ranges.

**Table S7.** Endpoint-normalization timing summary

| Endpoint interval | Optical-flow median $T_{20–80}$ (s) | NCC median $T_{20–80}$ (s) | Status |
|---|---|---|---|
| 10.0–10.5 s | 0.867 | 0.910 | Primary grid |
| 10.1–10.6 s | 0.866 | 0.912 | Primary grid |
| 10.2–10.7 s | 0.866 | 0.912 | Primary grid |
| 10.3–10.8 s | 0.882 | 0.917 | Primary grid |
| 10.4–10.9 s | 0.894 | 0.923 | Primary grid |
| 10.5–11.0 s | 0.901 | 0.928 | Primary grid |
| 14.033–14.533 s | 0.972 | 0.998 | Exploratory late |

Note. The six primary rows report method medians for every prespecified interval. $V_{20\text{-}80} = 0.829759$ m/$T_{20\text{-}80}$, so the endpoint pixel vector cancels exactly; rates are calculated from unrounded $T_{20\text{-}80}$ values. The field team's approximate

±0.05 m uncertainty for each measured component corresponds to about ±3.6% on $D_f$ and calibrated rate and is reported separately.

**Table S8.** Geometry-restricted feature counts and viability

| Analysis | Near-side reference candidates / accepted | Far-side target candidates / accepted | Outcome |
|---|---|---|---|
| Optical flow | 29 / 29 | 16 / 16 | 181/181 valid frames |
| NCC, 31 px | 4 / 2 | 2 / 1 | Insufficient for model fit |
| NCC, 41 px | 4 / 2 | 2 / 1 | Insufficient for model fit |
| NCC, 51 px | 4 / 0 | 2 / 0 | Insufficient for model fit |

Note. Counts were produced with the documented selection and containment rules after the windows were frozen. All other scene features were geometrically excluded.

**Table S9.** Outcome of methodologically distinct validation and artifact diagnostics

| Diagnostic | Main quantitative result | Interpretation |
|---|---|---|
| Geometry-restricted optical flow | Similarity 0.826–0.852 s; affine 0.842–0.892 s | Models overlap and recover the rapid rise |
| Whole-window NCC | 0.826 / 0.846 s; endpoint 24.868 px | Reproduces scale; timing is shorter than primary NCC |
| Late normalization | 0.972 / 0.998 s | Original interval not a settled plateau |
| Reference-only homography | OF retains 58%; NCC −4%; held-out fail | Inconclusive / mixed |
| Display border | Horizontal and late-tracking gates fail | Non-identifying |
| Display-plane direct alignment | Controlled real-bezel recovery gates fail | No projective correction applied to earthquake history |

Note. Failed and inconclusive tests are retained to prevent selective reporting. None establishes physical overshoot, and the diagnostic timing values are not a complete uncertainty envelope.

**Table S10.** Sensitivity to the common-motion model in the primary geometry

| Method | Similarity $T_{20–80}$ range (s) | Affine $T_{20–80}$ range (s) | Paired affine-minus-similarity difference (s) |
|---|---|---|---|
| Optical flow | 0.865–0.902 | 0.869–0.901 | −0.003 to +0.005 (median +0.002) |
| NCC | 0.880–0.893 | 0.941–0.948 | +0.055 to +0.064 (median +0.060) |

Note. Ranges are across the six normalization intervals. Paired differences compare matched intervals. The 0.068-s NCC common-motion envelope cited in the main text spans both model clusters and is not the paired median model gap.

## Supporting Figures

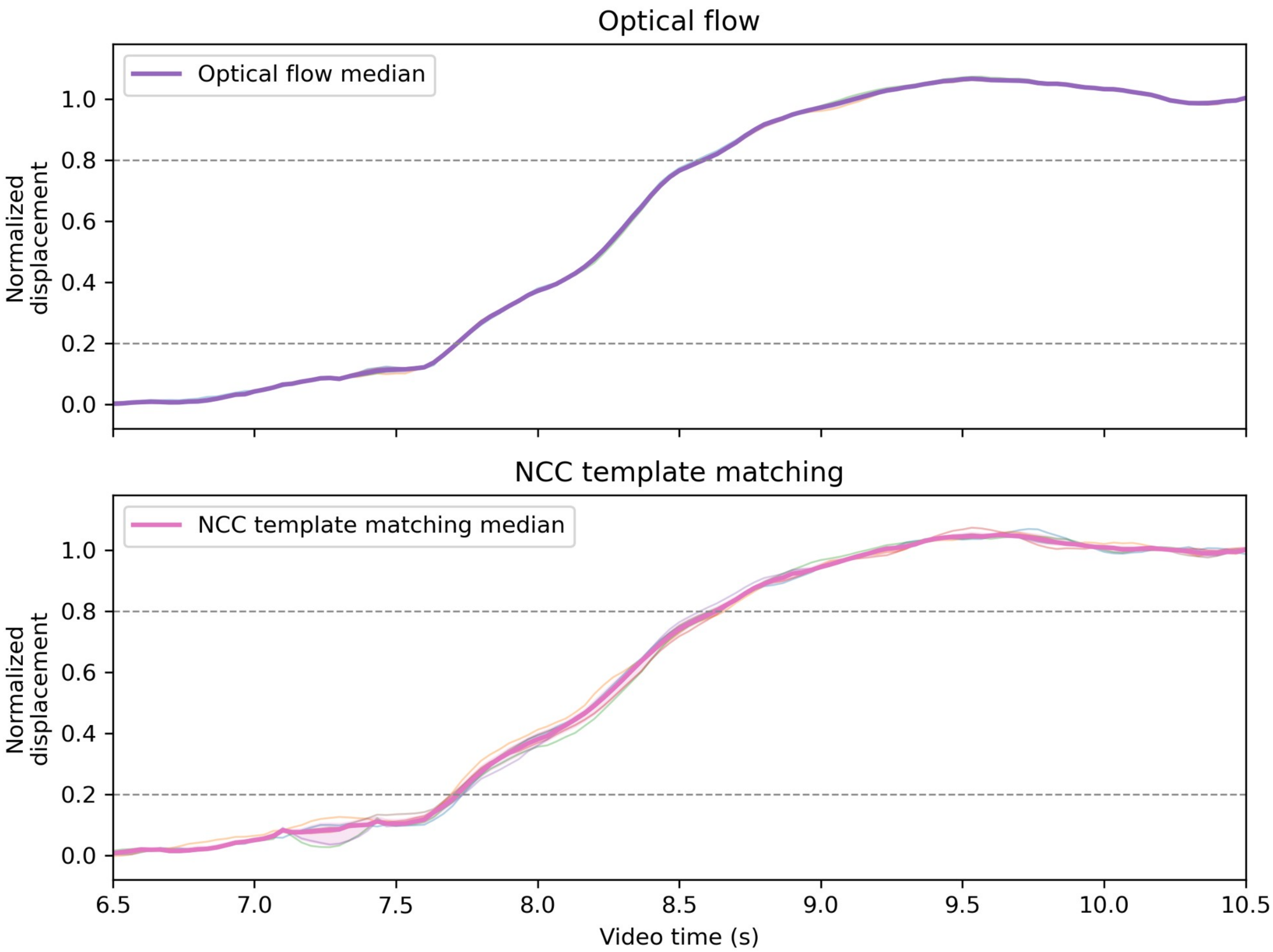

**Figure S1.** Normalized displacement histories for the fixed optical-flow and NCC tracker ensembles at 11-frame smoothing. Thin curves are individual variants; thick curves are method medians. Displacement is normalized to the original 10.0–10.5-s interval, not to a demonstrated settled plateau. Dashed lines mark 20% and 80%.

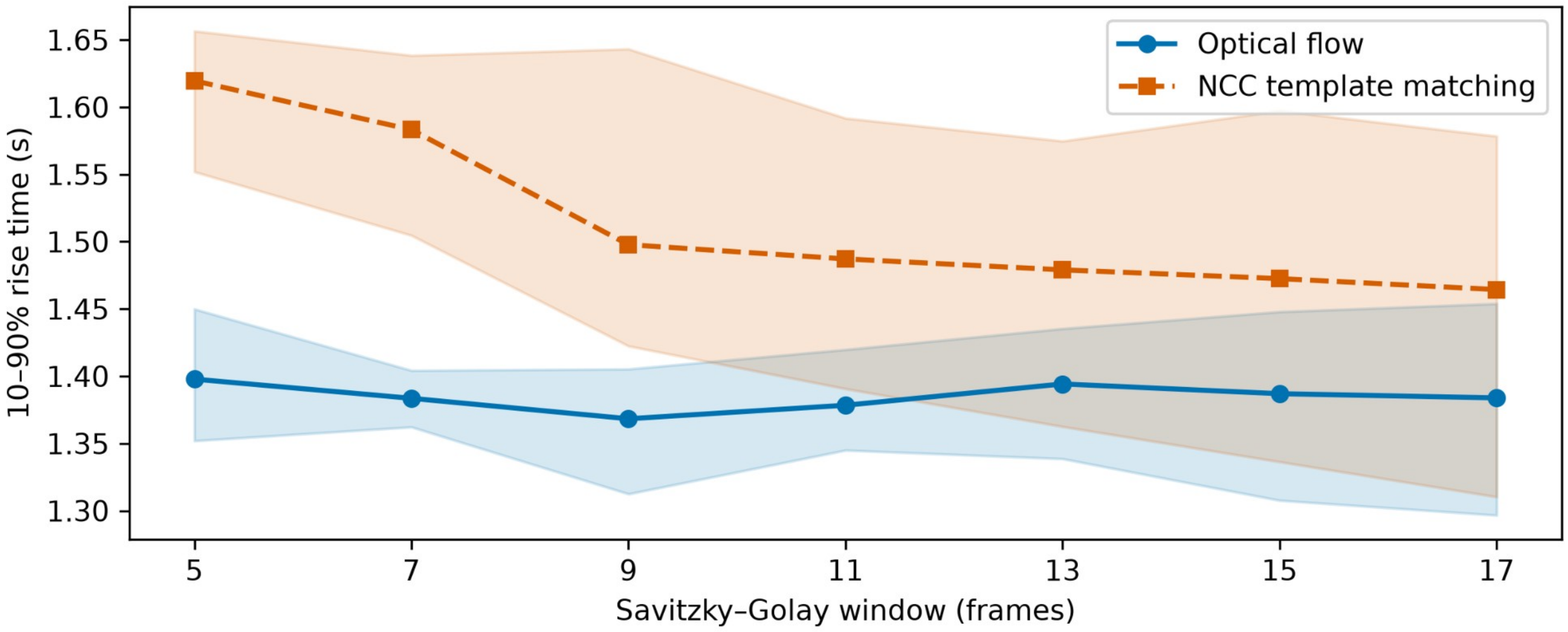

Figure S2. Sensitivity of the secondary 10–90% rise time to Savitzky–Golay smoothing-window length. Optical flow is shown by a solid line with circles and NCC by a dashed line with squares. Curves show method medians, and shaded bands indicate the 16th–84th percentile processing spread across the fixed final variants at each smoothing window. The primary 20–80% smoothing sensitivity is shown in main Figure 4a.

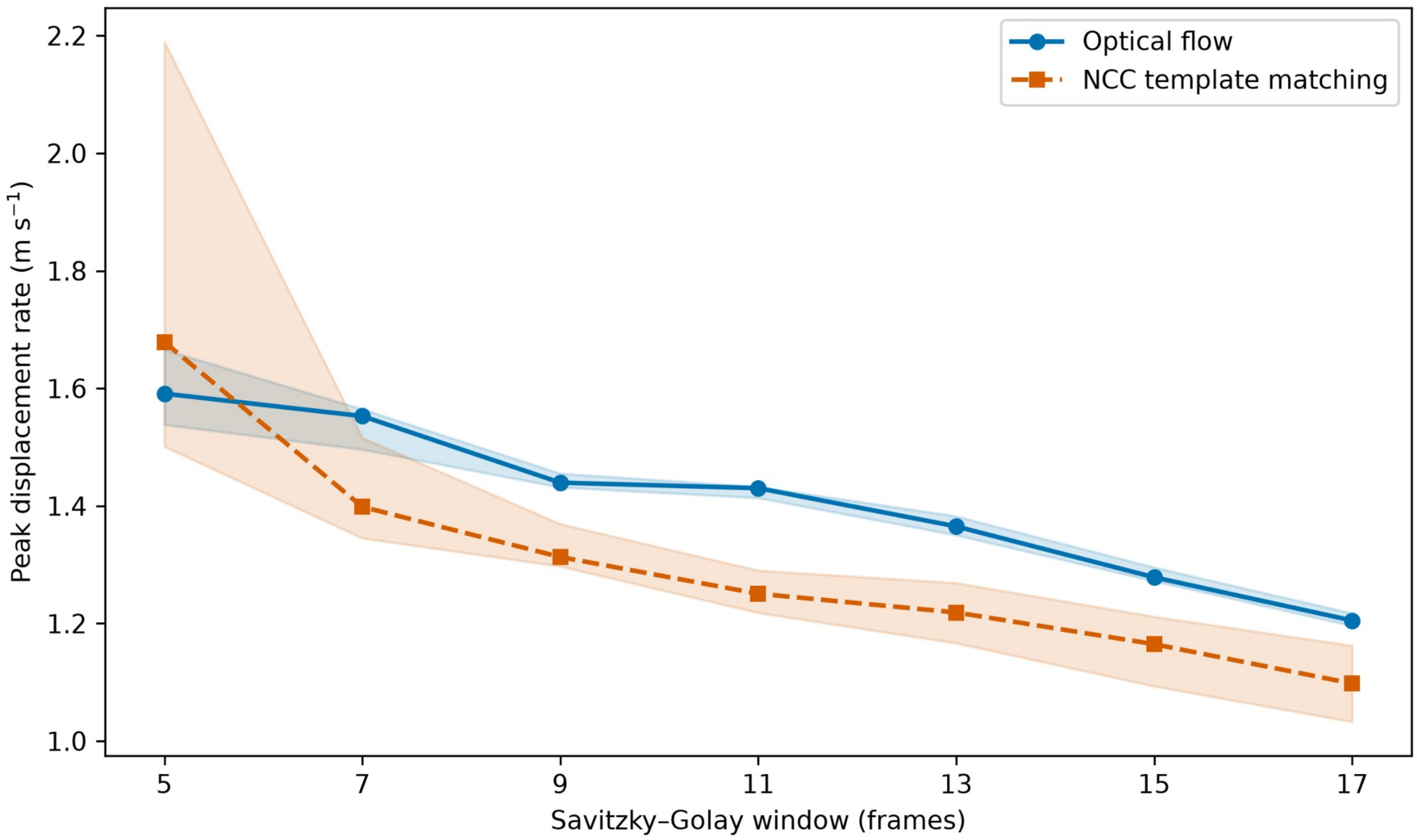

**Figure S3.** Sensitivity of peak displacement rate to Savitzky–Golay smoothing-window length. Optical flow is shown by a solid line with circles and NCC by a dashed line with squares. Curves show method medians, and shaded bands indicate the 16th–84th percentile processing spread across the fixed final variants at each smoothing window. Peak-rate amplitude decreases with increasing smoothing and is particularly variable for NCC at the shortest window, motivating the use of the mean 20–80% displacement rate as the primary rate measure in the main text.

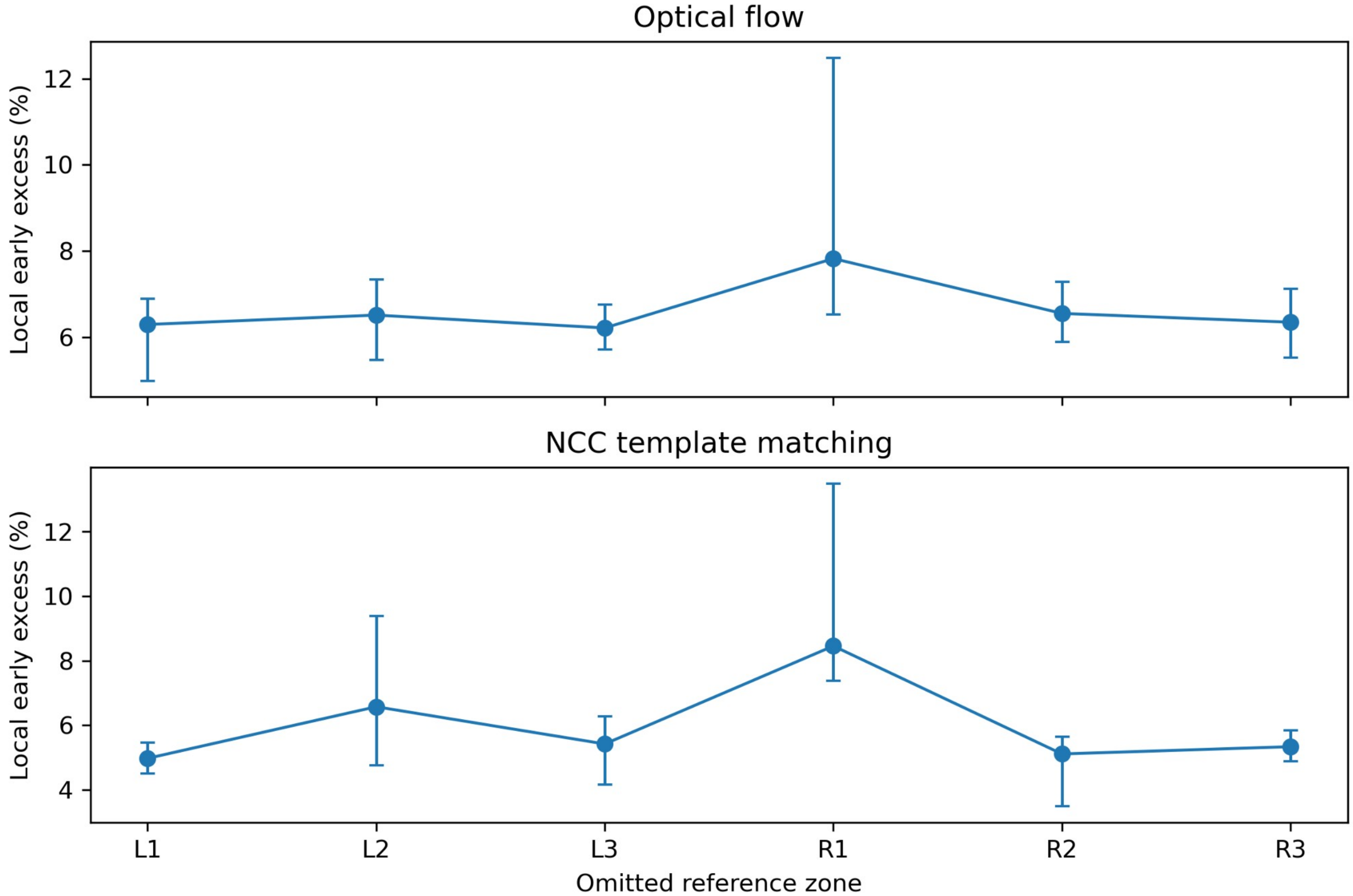

**Figure S4.** Sensitivity of the local early excess to omission of individual reference zones. The target is fixed while L1–L3 and R1–R3 are omitted one at a time. Points show medians and error bars show the 16th–84th percentile processing spread. The excess remains positive for every omission, so no single reference zone creates the early peak/decline pattern. This internal robustness test does not establish physical overshoot.

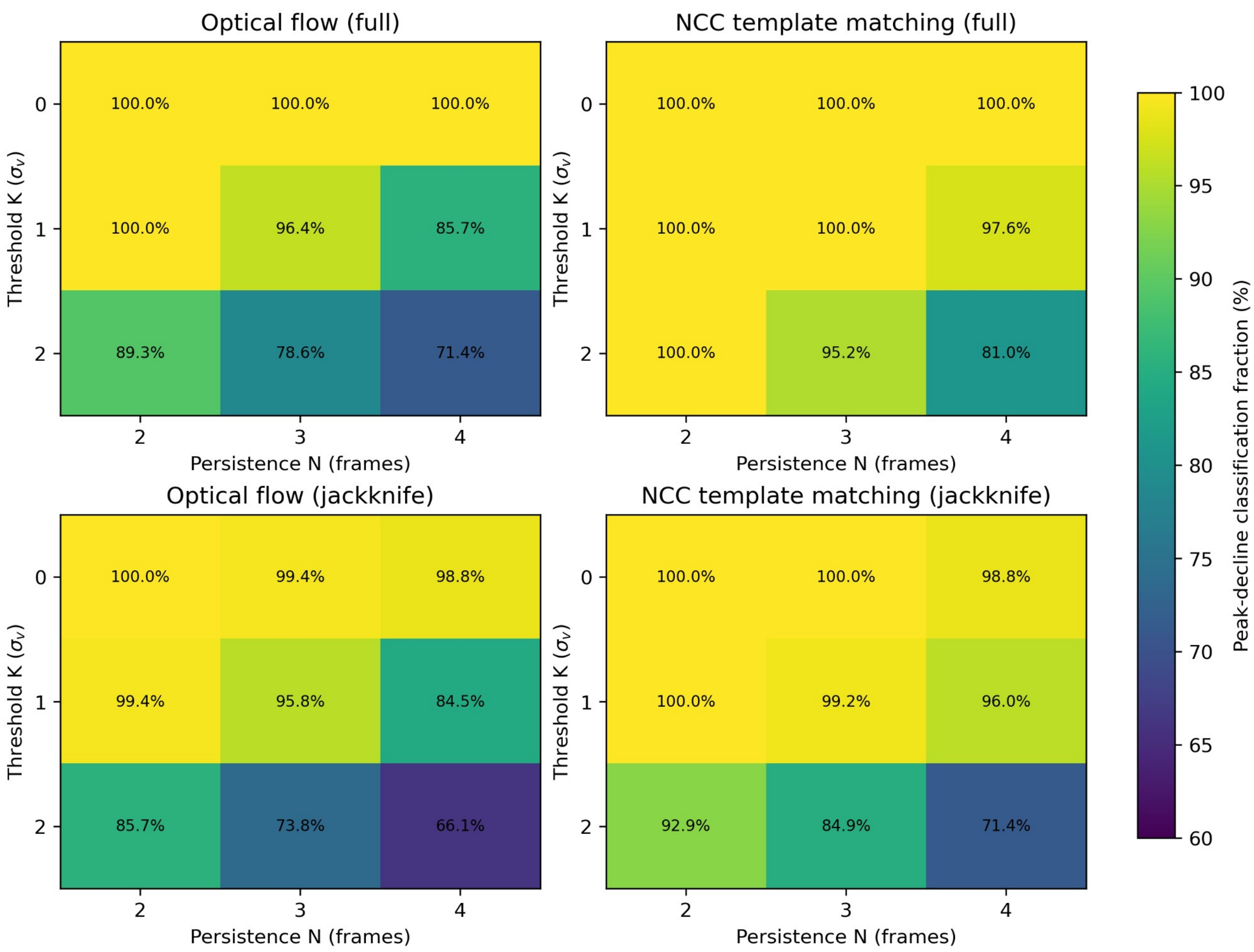

**Figure S5.** Sensitivity of the within-video early peak-and-decline classification to velocity threshold K and persistence N. Colors give the percentage satisfying the internal criterion for full-reference and jackknife analyses. The criterion identifies an early peak followed by declining apparent displacement under the original normalization; it does not establish a permanent endpoint or physical overshoot.

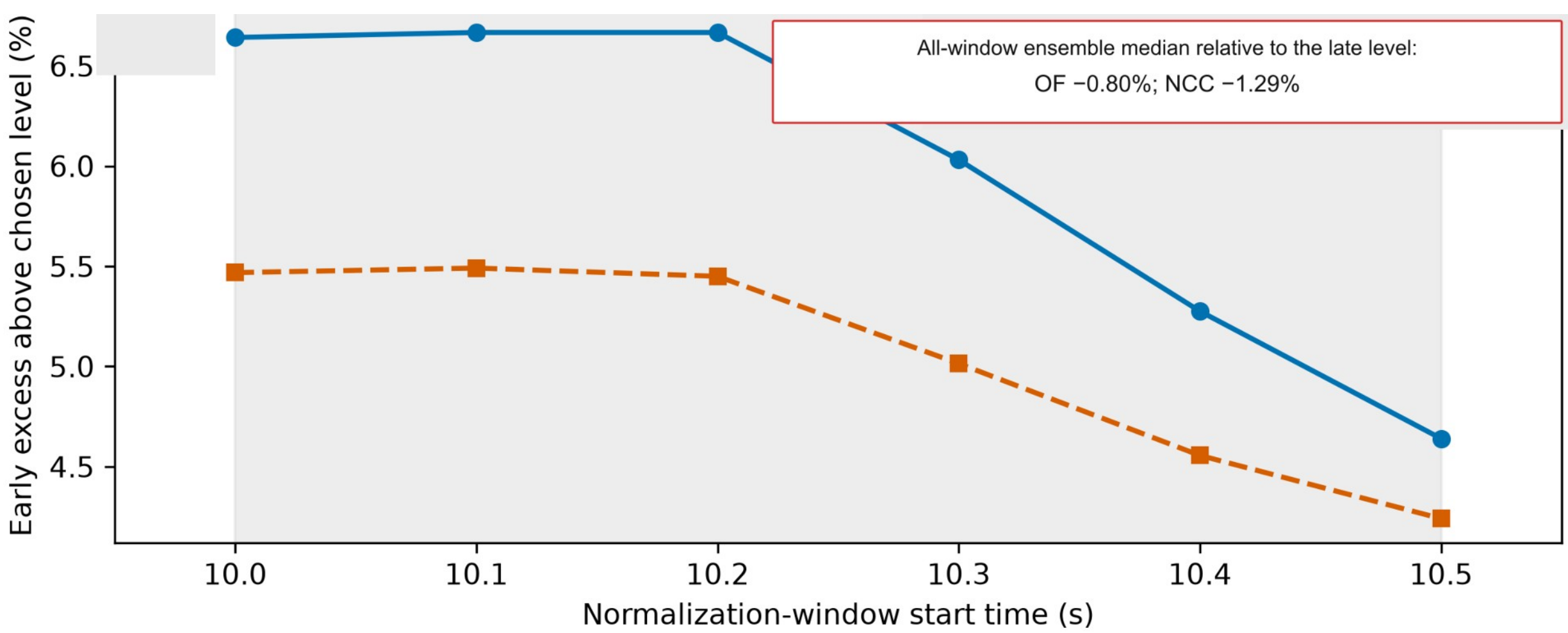

Figure S6. Local early excess above each of the six original 0.5-s normalization intervals. Optical flow is shown by a solid line with circles and NCC by a dashed line with squares. The red-bordered annotation gives the all-window ensemble median relative to the exploratory late level: −0.80% for optical flow and −1.29% for NCC. Within the primary grid, the excess varies by factors of 1.44 and 1.29, respectively, below the frozen 1.5× threshold; the exploratory higher late level reverses the sign. Timing sensitivity is shown in main Figure 4b.

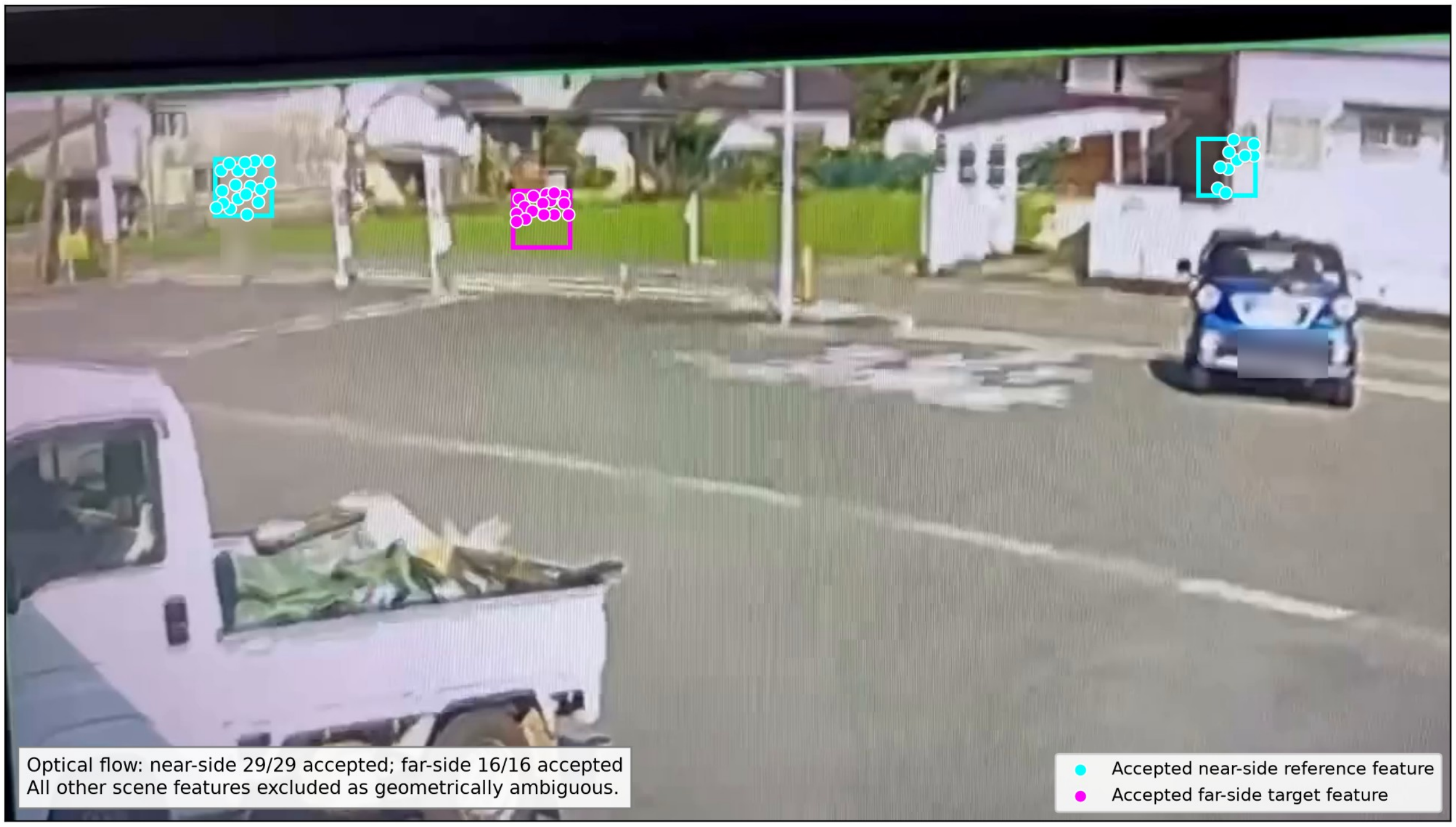

**Figure S7.** Exact Hirano windows and accepted optical-flow features on the 5.0-s pre-slip frame. Cyan marks the two near-side reference windows and magenta the far-side target window. Optical flow accepts 29 of 29 near-side candidates and 16 of 16 far-side candidates. All pixels outside the three exact 50 × 50-pixel windows were excluded from this geometry-restricted test. Privacy blurring affects display only. Video frame from the public Threads post cited in the text; excluded from the article's CC BY 4.0 licence.

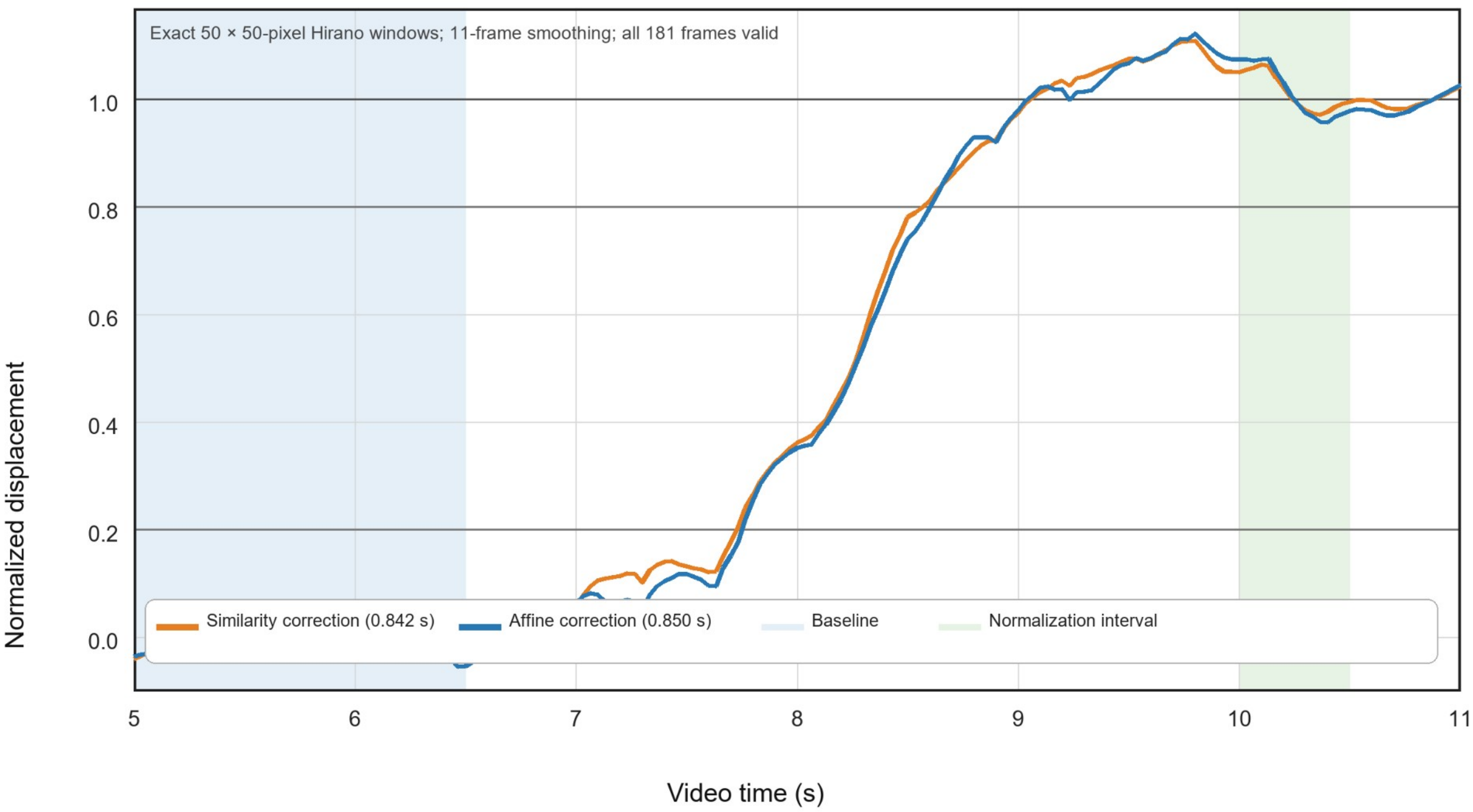

**Figure S8.** Geometry-restricted optical-flow histories at 11-frame smoothing using Hirano's exact windows. Similarity and affine common-motion correction give $T_{20-80}$ = 0.842 and 0.850 s, respectively, and both recover the same rapid main rise. The shaded intervals mark the 5.0–6.5-s baseline and 10.0–10.5-s normalization interval.

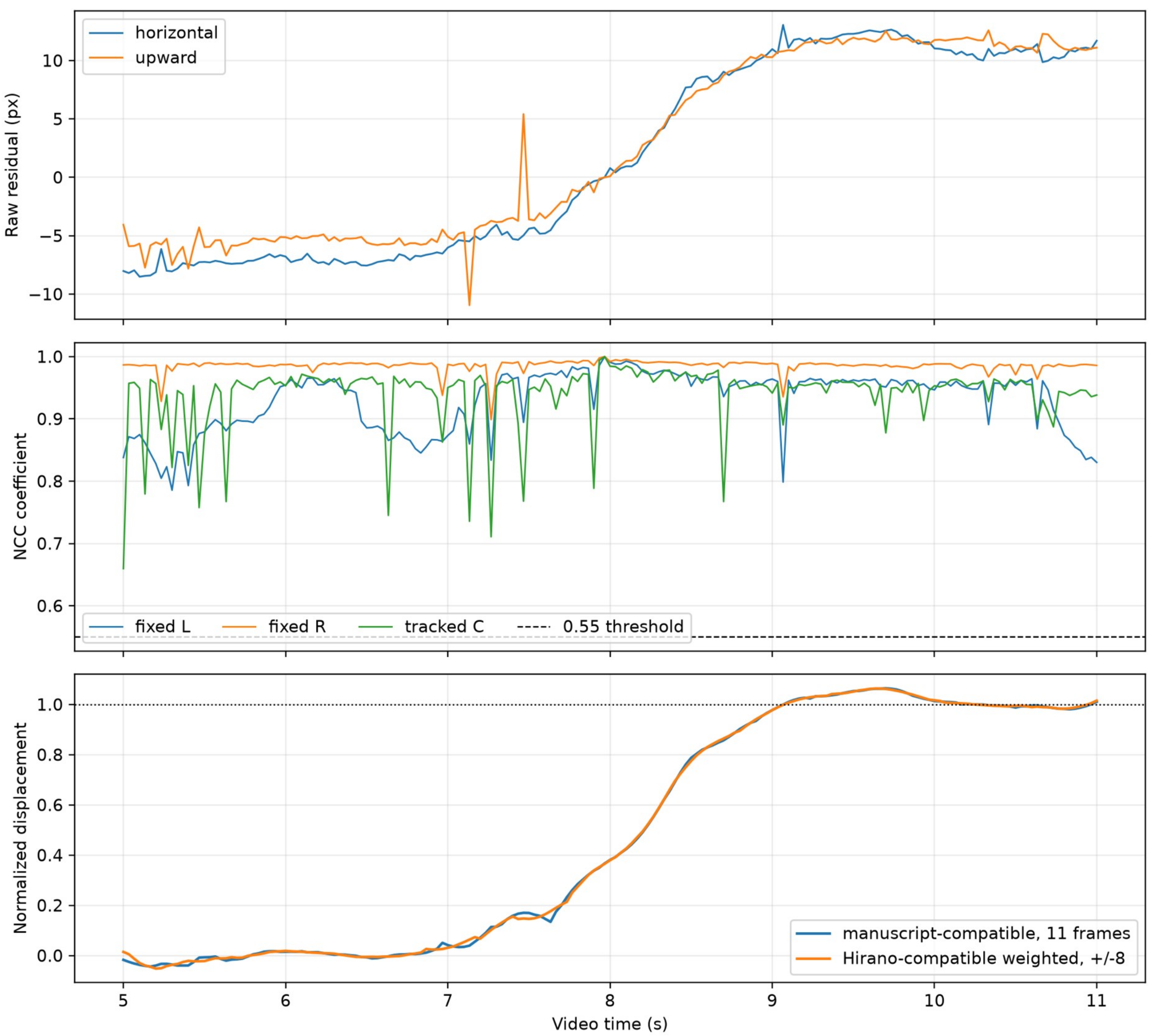

**Figure S9.** Corrected whole-window NCC cross-check using Hirano's exact central-frame templates. Top: raw horizontal and upward residuals. Middle: NCC coefficients for the two fixed windows and tracked target window. Bottom: normalized histories for manuscript-compatible and Hirano-compatible processing. The 24.868-pixel endpoint agrees within 0.11% with exact-window optical flow and 0.10% with primary NCC. $T_{20\text{–}80}$ is 0.826 and 0.846 s. The rise is rapid but shorter than primary NCC, local-window direction is less stable than magnitude, and the frozen early-excess significance gate is not passed.

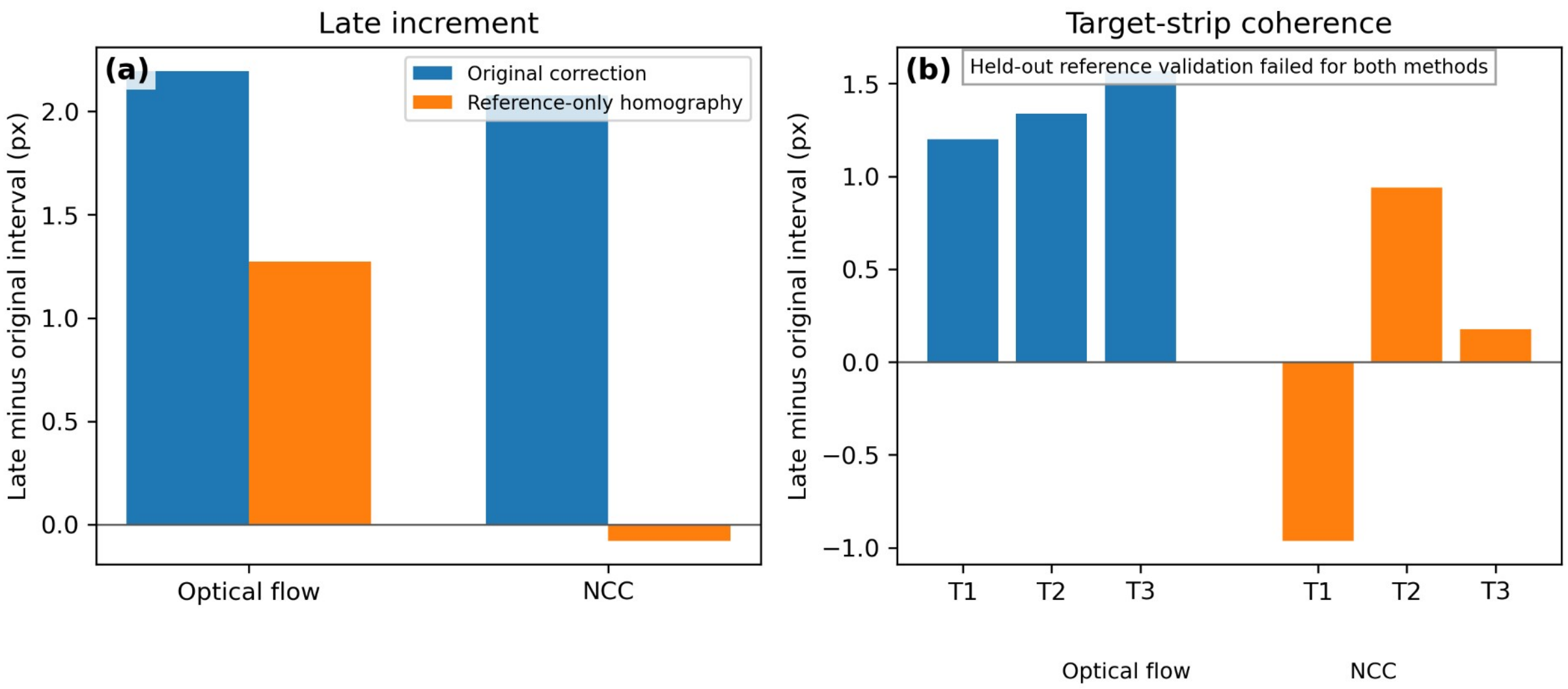

**Figure S10.** Reference-only projective diagnostic. (a) Late increment before and after homography correction. (b) Late increments in three adjacent left-to-right strips (T1–T3) of the primary target ROI. Optical flow retains a coherent positive increment, whereas NCC becomes spatially incoherent; held-out reference validation fails for both. The frozen outcome is inconclusive/mixed.

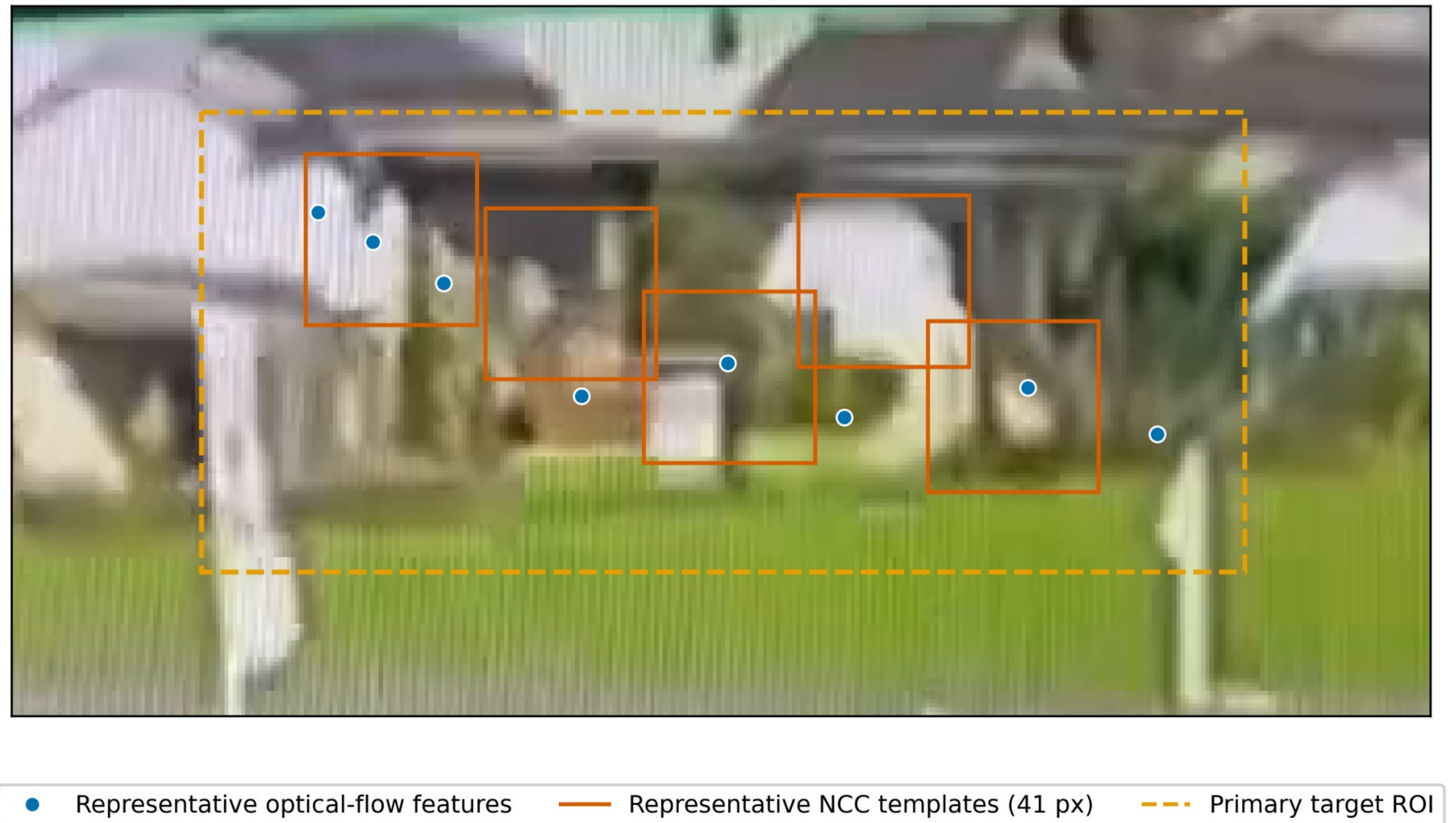

**Figure S11.** Detailed view of the primary far-side target ROI on the 5.0-s pre-slip frame. Blue circles show representative optical-flow features and orange squares show representative 41-pixel NCC templates; the dashed orange outline is the primary target ROI. Trackable features and templates are concentrated on the upper, distant structures because the lower paddy surface has comparatively little stable texture. The displayed locations are representative annotations of the frozen tracking setup, not a new feature-selection analysis. Video frame from the public Threads post cited in the text; excluded from the article's CC BY 4.0 licence.

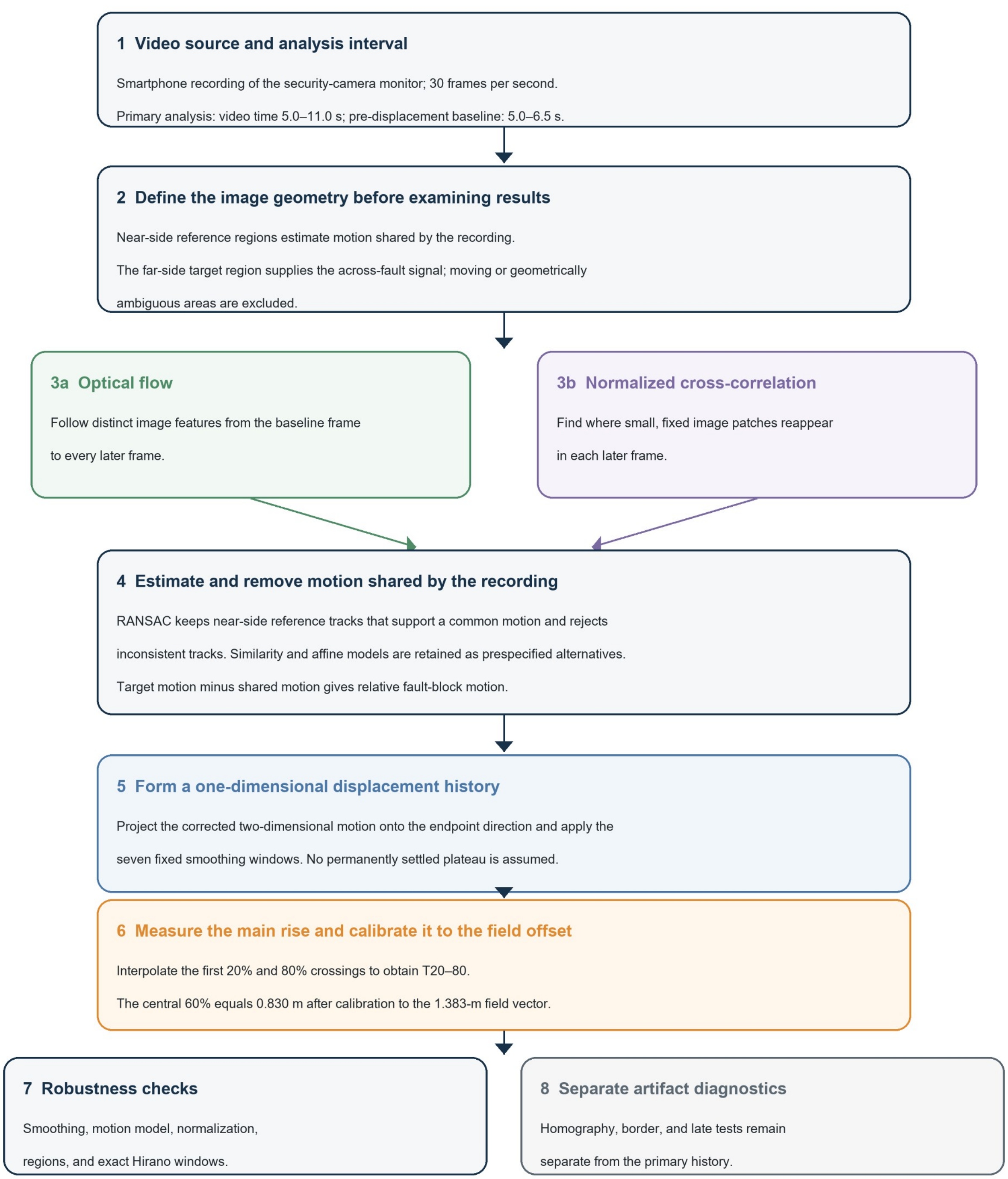

Figure S12. Overview of the displacement-analysis workflow. Image regions were defined before results were examined. Two independent trackers measured image motion. Near-side reference tracks were then used to estimate motion shared by the recording, and subtracting that motion from the far-side target tracks yielded relative displacement across the fault. Prespecified projection, smoothing, crossing-time measurement, and field calibration produced the primary approximately 0.9-s result. Robustness checks and secondary-recording artifact diagnostics were retained as separate branches; the latter were not used to modify the primary displacement history.